\documentclass[12pt]{article}

\usepackage[margin=1in]{geometry}
\usepackage{setspace}
\usepackage{amsmath,amssymb,bm}
\usepackage{graphicx}
\usepackage{cite}
\usepackage{hyperref}

\begin{document}

\title{Power-Law Relaxation of Non-Gaussian Parameter and Self-Dynamic Structure Factor in Multidimensional Rugged Energy Landscapes}

\author{Biman Bagchi\\
Solid State and Structural Chemistry Unit\\
Indian Institute of Science, Bengaluru 560012, India}

\date{\today}

\maketitle

\begin{abstract}

Ruggedness of the underlying energy landscape gives rise to heterogeneous mobility and non-Gaussian diffusion. We develop a theoretical framework for tagged-particle diffusion in multidimensional rugged energy landscapes modeled as correlated quenched Gaussian random fields. Using the self-propagator and self-dynamic structure factor, we characterize finite-time diffusion beyond the effective diffusion coefficient. We determine the effects of dimensionality, spatial correlations, and initial preparation. By introducing a coarse-grained mobility field and a mobility-memory approximation, we relate the non-Gaussian parameter to the time correlation of the mobility sampled by the particle. In the homogenized diffusive regime, the mobility correlation decays algebraically, leading to long-time relaxation of the non-Gaussian parameter as $t^{-1/2}$ in one dimension, $(\ln t)/t$ in two dimensions, and $t^{-1}$ for $d>2$, with amplitudes that depend on dimensionality and the initial ensemble. Our results show that rugged energy landscapes leave distinct signatures in the effective diffusion coefficient, self-dynamic structure factor, and relaxation of non-Gaussian fluctuations.

\end{abstract}

\section{Introduction}

Diffusion in rugged energy landscapes is a central problem in chemical physics, biophysics, condensed matter physics, and the theory of disordered systems. A diffusing particle, polymer segment, protein coordinate, or biomolecular complex often moves through an effective free-energy surface containing many small wells and barriers. These local energetic irregularities may arise from structural disorder, conformational heterogeneity, solvent fluctuations, sequence-dependent interactions, or microscopic packing constraints. Even when the underlying microscopic motion is locally Brownian, energetic ruggedness can strongly reduce the observed long-time transport coefficient.

In fact, rugged- or rough-landscape descriptions have been used in many related settings, including diffusion in periodic or rough potentials, glassy and inherent-structure dynamics, protein-folding landscapes, biomolecular search on sequence-dependent energy landscapes, and heterogeneous non-Gaussian transport \cite{LifsonJackson1962,Keyes1997,LiKeyes1999,Onuchic1997,SlutskyMirny2004,Chechkin2017,PachecoPozo2023,BagchiBlaineyXie2008,Blainey2009}.

The same problem has also been studied from the viewpoint of disordered transport theory. Random-barrier and random-trap models, continuous-time random walks, effective-medium theories, and random walks on rugged potential surfaces have all shown that transport in a disordered landscape can differ qualitatively from ordinary diffusion, especially when the disorder is strong or the dimensionality is low \cite{SteinNewman1995,Dyre1988,DyreJacobsen1995,Chechkin2017,PachecoPozo2023}. These studies provide an important theoretical background for the present work, but most of them focus either on effective transport coefficients, anomalous mean-square displacements, or heterogeneous diffusivity models. Our objective here is complementary: to formulate the problem at the level of the full propagator and to obtain the dimension-dependent power-law relaxation of the non-Gaussian parameter.

A landmark contribution to this problem was made by Zwanzig, who considered diffusion in a one-dimensional rough potential consisting of a smooth background part and a rapidly varying rough component.\cite{Zwanzig1988} By analyzing the mean first passage time and locally averaging over the roughness, Zwanzig obtained a remarkably simple expression for the reduction of the diffusion coefficient by Gaussian energetic disorder. This result has been widely influential because it suggests that the complicated microscopic structure of a rugged landscape can, under suitable conditions, be represented by a single measure of the roughness amplitude.

The possibility that low-dimensional rugged landscapes can display pathological dynamics was emphasized by Stein and Newman in their study of broken ergodicity and the geometry of rugged landscapes \cite{SteinNewman1995}. Using random-walk models corresponding to diffusion on random potential surfaces, they showed that motion in a rugged landscape can be organized by drainage-like pathways, basins, and barriers. This hydrological picture makes clear why low-dimensional diffusion is special: a walker may be forced to revisit the same valleys and bottlenecks rather than bypass them through alternate directions. Thus one-dimensional and low-dimensional rugged-landscape diffusion should not be viewed merely as a simplified version of high-dimensional transport, but as a setting where geometry can qualitatively constrain the dynamics.

The assumptions behind Zwanzig's result are subtle. His treatment is continuous and one-dimensional. It does not require the introduction of hopping rates because the dynamics is formulated at the Smoluchowski level. The roughness is removed by a local spatial averaging procedure, and the final result is based on the mean first passage time rather than on a full time-dependent propagator. Thus, the theory does not explicitly address how the propagator depends on the initial distribution of the particle, nor does it provide a direct treatment of arbitrary dimensionality.

Banerjee, Biswas, Seki, and Bagchi (BBSB) revisited this problem by studying diffusion on a discrete one-dimensional Gaussian site-energy landscape. \cite{BBSB2014} In a discrete landscape, the site energies alone do not define the dynamics; one must also specify transition rates between neighboring sites.\cite{BBSB2014} For nearest-neighbor activated hopping satisfying detailed balance, their exact mean-first-passage-time analysis produced an additional error-function correction to Zwanzig's exponential result. The physical origin of this correction is the presence of rare but long-lived local configurations in which a deep minimum is flanked by two high-energy neighboring sites. These three-site traps are especially important in one dimension because the walker cannot bypass them. Thus, the deviation from Zwanzig's result is not merely a small correction; it reflects a qualitative rare-event effect produced by the local structure of an uncorrelated discrete landscape.

Spatial correlations change this conclusion. \cite{SBB2016,Bagchi2026}  If neighboring site energies are correlated, sharp high-low-high configurations are suppressed. In the one-dimensional correlated landscape, the error-function correction is weakened as the correlation length increases and disappears in the strongly correlated limit. In that limit Zwanzig's result is recovered. This provides a clear physical interpretation of the role of correlations: spatial correlations smooth out local roughness increments and suppress the rare traps that dominate the uncorrelated discrete problem.

The role of dimensionality was subsequently examined by Seki, Bagchi, and Bagchi.\cite{SBB2016} Their work showed that diffusion in a rugged landscape is anomalously sensitive to dimensionality, especially in going from one to two dimensions. In one dimension, the walker repeatedly retraces its path when it encounters a large barrier or a deep local trap. In two and three dimensions, alternate pathways allow the walker to partially avoid such bottlenecks. Using effective-medium theory, Seki, Bagchi, and Bagchi showed that the diffusion coefficient can increase strongly from one to two dimensions, whereas the difference between higher successive dimensions is much weaker. Their analysis clarified why one-dimensional rugged-landscape diffusion is special. However, the central object of that work remained the long-time effective diffusion constant, not the full propagator.

The present work is motivated by the need to go beyond the asymptotic diffusion coefficient. The effective long-time diffusion constant is an important quantity, but it contains only limited information. It does not tell us how the system approaches the diffusive limit, how long the memory of the initial condition persists, how trapping affects the displacement distribution at intermediate times, or how non-Gaussian the propagator is before the central-limit regime is reached. These questions become especially important in rugged landscapes, where rare traps and spatial correlations can generate broad distributions of residence times and strongly heterogeneous transport.

We therefore take the propagator as the primary object. For a fixed quenched landscape, the probability distribution evolves under a transition-rate operator determined by that realization of the landscape. The displacement propagator can then be constructed for a specified initial distribution and subsequently averaged over the ensemble of rugged landscapes. This formulation makes the role of initial conditions explicit. Neither Zwanzig's original treatment nor the BBSB mean-first-passage-time analysis required an arbitrary initial occupation distribution, because both focused on the long-time effective diffusion coefficient. In contrast, the propagator depends on how the particle is initially prepared. Even when a well-defined asymptotic diffusion coefficient exists and becomes independent of the initial distribution, the finite-time propagator and its cumulants can retain strong memory of the initial preparation.

A useful feature of the propagator formulation is that it permits controlled variation of the initial ensemble. In many theoretical treatments of rugged landscapes one implicitly assumes equilibrium preparation, or considers only the long-time diffusion coefficient for which the initial state is eventually forgotten. However, experiments and simulations often probe finite-time dynamics, where the initial distribution can matter. A particle may be placed uniformly in the sample, thermally equilibrated in the energy landscape, preferentially initialized in low-energy regions, or selectively excited into high-energy regions. We therefore introduce a preparation parameter, denoted below by $\beta_0$, which allows these different initial ensembles to be described within a single framework. This parameter is not a new measure of ruggedness; it specifies how the initial condition samples a rugged landscape whose energetic width is fixed.

A particularly important quantity is the non-Gaussian parameter. For a Gaussian displacement distribution this quantity vanishes, whereas a nonzero value signals a fourth-order deviation from Gaussian diffusion. It is therefore sensitive to intermittent trapping, heterogeneous mobility, and incomplete sampling of the rugged landscape. In the present problem, however, $\alpha_2(t|\rho_0)$ is not a universal function of time independent of preparation. At short and intermediate times it may contain substantial contributions from the way the initial ensemble samples the quenched landscape. Thus equilibrium, uniform, trap-biased, and high-energy-biased preparations can give different finite-time non-Gaussian responses. While the diffusion constant is controlled by the second moment in the long-time limit, the non-Gaussian parameter requires the fourth moment, or equivalently the relevant fourth-order cumulant combination, of the propagator. A systematic propagator-based treatment of this quantity in dimensional correlated rugged landscapes has not been developed within the Zwanzig--BBSB--SBB framework.

A more complete treatment of non-Gaussian diffusion also requires two related quantities that are often implicit in discussions of rugged landscapes. The first is the correlation function of the mobility sampled by the particle. In a strict microscopic sense, a unique local diffusivity is not defined in a quenched hopping landscape; it becomes meaningful only as a coarse-grained mobility field specified on an appropriate length and time scale. Within such a coarse-grained description, the spatial correlation of mobility fluctuations and the corresponding sampled temporal correlation describe how mobility heterogeneity persists along the trajectory. The second quantity is the self-dynamic structure factor, $F_s(k,t|\rho_0)=\langle \exp[i{\bf k}\cdot\Delta{\bf r}(t)]\rangle_{\rho_0}$, which is the experimentally accessible Fourier transform of the self-propagator. Its small-$k$ cumulant expansion naturally brings in the fourth-order displacement cumulant and therefore the non-Gaussian parameter $\alpha_2(t|\rho_0)$. These two objects provide the bridge between coarse-grained mobility memory and measurable deviations from Gaussian diffusion.

A key outcome of the present formulation is that dimensionality can affect not only the long-time diffusion coefficient but also the relaxation of non-Gaussianity. This statement is conditional on the existence of a homogenized diffusive regime: the effective diffusion coefficient $D_{\rm eff}$ must exist, the quenched disorder must be self-averaging, and the central part of the self-propagator must have ordinary diffusive scaling at long times. Under this assumption, the mobility-memory approximation developed below converts a finite-range spatial mobility correlation into a sampled temporal memory. For a Gaussian-diffusive central propagator and a mobility correlation length $\xi_D$, this memory contains the factor $[1+2D_{\rm eff}t/\xi_D^2]^{-d/2}$. The resulting long-time mobility-induced contribution to $\alpha_2(t|\rho_0)$ has different asymptotic forms in one, two, and higher dimensions. This provides a propagator-level counterpart of the anomalous dimensionality dependence found by Seki, Bagchi, and Bagchi for the effective diffusion coefficient, while making explicit the additional assumptions required at the level of the full propagator.

Throughout this work, $D_{\rm eff}$ denotes the asymptotic long-time diffusion coefficient when such a limit exists. For a $d$-dimensional system, it is obtained from the long-time limit of the mean-square displacement, or equivalently from the long-time limit of the apparent diffusion coefficient $D_{\rm app}(t)$. Thus $D_{\rm eff}$ measures the final coarse-grained transport rate after averaging over the rugged landscape. It should be distinguished from the bare diffusion coefficient $D_0$, which describes motion in the absence of energetic ruggedness, from the finite-time diffusion tensor or apparent diffusion coefficient, which may depend on $\rho_0$, and from the full propagator, which contains finite-time and initial-state-dependent information lost in the long-time limit.

The objective of this paper is to develop such a formulation. We first review the one-dimensional Zwanzig and BBSB results, emphasizing the difference between continuous rough-potential diffusion and discrete site-energy hopping, the origin of the error-function correction, the role of spatial correlations, and the absence of explicit initial-condition dependence in the earlier long-time treatments. We then summarize the multidimensional effective-medium analysis of Seki, Bagchi, and Bagchi, focusing on the anomalous role of one dimension and the physical importance of bypass pathways in higher dimensions. Finally, we introduce a dimensional propagator framework for correlated rugged landscapes and use it to define time-dependent observables that go beyond the effective diffusion coefficient, especially the non-Gaussian parameter. The resulting asymptotic predictions should be understood as conditional results for the homogenized diffusive regime, rather than as exact statements for arbitrary quenched landscapes. This approach also provides a natural foundation for future extensions to dynamically fluctuating energy landscapes, where the ruggedness itself evolves in time.

A further motivation for the present formulation is experimental observability. The long-time diffusion coefficient provides one measure of ruggedness because it quantifies the departure from the bare diffusion coefficient $D_0$. Such an effective coefficient may be inferred from transport measurements, including conductivity in ionic or charge-transport systems, or from the long-time slope of the mean-square displacement in simulations and single-particle experiments. However, a single diffusion coefficient does not reveal whether the dynamics is Gaussian, whether different particles experience different local mobilities, or whether the system retains memory of its initial preparation. For this reason, the propagator and its higher moments are more informative. In particular, the non-Gaussian parameter $\alpha_2(t|\rho_0)$, which involves the fourth moment of the displacement distribution, provides an experimentally accessible measure of dynamic heterogeneity, transient trapping, and preparation dependence. It can be obtained from optical particle-tracking experiments, fluorescence-based tracer measurements, scattering measurements through the self-intermediate scattering function, and numerical trajectories. Thus, the present work is concerned not only with the renormalization of the diffusion constant by ruggedness, but also with experimentally measurable deviations from Gaussian diffusion in finite-time propagator-level observables.

We have included two Appendices. Appendix A collects the important symbols and terminologies, as the paper has become rather involved. Appendix B presents a more involved analysis of the approach to the Gaussianity in this quenched rough landscape where non-Gaussianity is more of a rule than an exception.

%=============================== Section 2  ========================================
\section{One-Dimensional Rough Landscapes: Zwanzig, BBSB, and the Role of Gaussian Spatial Correlations}
%==================================================================================

Zwanzig considered diffusion in a one-dimensional rough potential written as
\begin{equation}
U(x)=U_0(x)+U_1(x),
\label{eq:rough_potential}
\end{equation}
where \(U_0(x)\) is a smooth background potential and \(U_1(x)\) is a rapidly varying roughness contribution.  The dynamics was formulated at the continuous Smoluchowski level, not as a hopping process on a discrete lattice.  This is an important distinction.  In Zwanzig's treatment, one does not need to introduce transition rates between sites; the effects of the potential enter through the Smoluchowski equation and through the mean first passage time.

By locally averaging over the roughness, Zwanzig obtained the effective diffusion coefficient
\begin{equation}
D_{\rm Z}
=
\frac{D_0}
{\left\langle e^{\beta U_1}\right\rangle
 \left\langle e^{-\beta U_1}\right\rangle},
\label{eq:Zwanzig_general}
\end{equation}
where \(D_0\) is the diffusion coefficient on the smooth potential and \(\beta=(k_{\rm B}T)^{-1}\).  For a Gaussian roughness distribution with zero mean and variance \(\epsilon^2\), Eq.~\eqref{eq:Zwanzig_general} gives
\begin{equation}
D_{\rm Z}
=
D_0\exp\left[-\beta^2\epsilon^2\right].
\label{eq:Zwanzig}
\end{equation}
The simplicity of Eq.~\eqref{eq:Zwanzig} has made it influential.  It suggests that the complicated microscopic structure of the rugged landscape can be absorbed into a single disorder variance.

However, the assumptions behind Eq.~\eqref{eq:Zwanzig} are subtle.  Zwanzig's treatment is continuous and one-dimensional.  The roughness is eliminated by a local spatial averaging procedure, and the final expression is based on the mean first passage time rather than on the full time-dependent propagator.  Thus, the theory does not explicitly address the dependence of the propagator on the initial distribution of the particle.  Nor does it provide, by itself, a direct formulation for arbitrary dimensionality.
%=================== BBSB  ===================
Banerjee, Biswas, Seki, and Bagchi revisited the problem by considering diffusion on a discrete one-dimensional Gaussian site-energy landscape.\cite{BBSB2014}  In this formulation, the energy is assigned to lattice sites.  Unlike in the continuous Smoluchowski description, the site energies alone do not define the dynamics.  One must also specify the transition rates between neighboring sites.  For nearest-neighbor activated hopping satisfying detailed balance, the transition rate from site \(i\) to a neighboring site \(j\) may be written as
\begin{equation}
\Gamma_{ij}
=
\begin{cases}
\Gamma_0, & U_j<U_i,\\
\Gamma_0\exp[-\beta(U_j-U_i)], & U_j\ge U_i.
\end{cases}
\label{eq:BBSB_rate}
\end{equation}
This rate rule preserves the site-energy character of the model, but it introduces a dynamical prescription that is absent from Zwanzig's continuous treatment.

For the uncorrelated Gaussian site-energy landscape, the exact one-dimensional mean-first-passage-time analysis gives
\begin{equation}
D_{\rm BBSB}
=
D_0\exp\left[-\beta^2\epsilon^2\right]
\left[
1+\operatorname{erf}\left(
\frac{\beta\epsilon}{2}
\right)
\right]^{-1}.
\label{eq:BBSB}
\end{equation}
The additional error-function factor in Eq.~\eqref{eq:BBSB} is absent in Zwanzig's expression, Eq.~\eqref{eq:Zwanzig}.  Its origin is the local ordering of neighboring site energies that enters the exact one-dimensional passage-time calculation.  Physically, it reflects rare but long-lived configurations in which a deep minimum is flanked by two high-energy neighboring sites.  These three-site traps are especially important in one dimension because the walker has no bypass route.  Thus, the deviation from Zwanzig's result is not merely a small correction; it reflects a qualitative rare-event effect produced by the local structure of an uncorrelated discrete landscape.

Spatial correlations change this conclusion.  Let the site energies obey a Gaussian correlation of the form
\begin{equation}
\left\langle U_i U_j \right\rangle
=
\epsilon^2
\exp\left[
-\frac{b^2(i-j)^2}{2\xi^2}
\right],
\label{eq:Gaussian_correlation}
\end{equation}
where \(b\) is the lattice spacing and \(\xi\) is the correlation length.  The relevant nearest-neighbor energy increment is then no longer controlled by the full variance \(\epsilon^2\), but by the reduced variance
\begin{equation}
\left\langle (U_{i+1}-U_i)^2 \right\rangle
=
2\epsilon^2
\left[
1-\exp\left(-\frac{b^2}{2\xi^2}\right)
\right].
\label{eq:increment_variance}
\end{equation}
The same conditional Gaussian average that produces the error-function correction in the uncorrelated case therefore gives a modified correction factor.  The one-dimensional correlated-landscape result is
\begin{equation}
D_{\rm corr}
=
D_0\exp\left[-\beta^2\epsilon^2\right]
\left[
1+\operatorname{erf}
\left(
\frac{\beta\epsilon}{2}
\sqrt{
1-\exp\left[-\frac{b^2}{2\xi^2}\right]
}
\right)
\right]^{-1}.
\label{eq:corr1d}
\end{equation}
In the uncorrelated limit, \(\xi\rightarrow 0\), one has
\begin{equation}
\exp\left[-\frac{b^2}{2\xi^2}\right]\rightarrow 0,
\label{eq:uncorr_limit}
\end{equation}
and Eq.~\eqref{eq:corr1d} reduces to Eq.~\eqref{eq:BBSB}.  In the strongly correlated limit, \(\xi\rightarrow\infty\), one has
\begin{equation}
1-\exp\left[-\frac{b^2}{2\xi^2}\right]\rightarrow 0,
\label{eq:corr_limit}
\end{equation}
so that the argument of the error function vanishes and Eq.~\eqref{eq:corr1d} reduces to Zwanzig's expression, Eq.~\eqref{eq:Zwanzig}.  This result gives a clear physical interpretation: Zwanzig's mean-field-like formula is recovered when spatial correlations smooth out the local roughness increments and suppress the rare traps that dominate the uncorrelated discrete problem.

The recent work of Bagchi on Gaussian spatial correlations returned to this issue from a broader perspective.  The central point was that the restoration of Zwanzig's result is not tied to a particular hopping rule.\cite{Bagchi2026}  Rather, it follows from the smoothing of roughness increments by spatial correlations.  In a continuous correlated Gaussian landscape, abrupt high-low-high configurations are strongly suppressed, and the local averaging implicit in Zwanzig's treatment becomes physically meaningful.  Thus, the correlated landscape provides a route back to Zwanzig's expression, while the uncorrelated discrete landscape exposes the rare-event correction hidden by the local smoothing procedure.

Three observations are important for the present work. First, Zwanzig's result belongs to a continuous one-dimensional Smoluchowski treatment with local spatial averaging.  Second, the BBSB correction belongs to a discrete one-dimensional site-energy landscape with activated nearest-neighbor hopping.  Third, Gaussian spatial correlations suppress the sharp local traps responsible for the error-function correction and thereby restore Zwanzig-like behavior in the strongly correlated limit.  None of these treatments, however, develops the full time-dependent propagator with explicit initial-condition dependence.  This is the point at which the present propagator-based approach begins.

%
%============================  Section 3  ============================
\section{Dimensionality Dependence: The Seki--Bagchi--Bagchi Effective-Medium Treatment}
%==========================================================

The one-dimensional results discussed in the preceding section show that local trapping configurations can strongly affect diffusion in rugged landscapes.  A natural question is whether this effect is special to one dimension or whether it persists in higher dimensions.  This question was addressed by Seki, Bagchi, and Bagchi, who studied diffusion in a \(d\)-dimensional rugged energy landscape and showed that the dependence on dimensionality is highly nontrivial.

The physical reason is simple.  In one dimension, a particle encountering a deep local trap or a large barrier has no alternate path.  It must either escape through one of the two neighboring sites or repeatedly return to the same region.  Thus, local traps act as serial bottlenecks.  In two and three dimensions, by contrast, the particle can partially avoid unfavorable regions by moving through alternate pathways.  Dimensionality therefore changes not only the number of neighboring sites but also the connectivity of the landscape.  This makes the transition from one to two dimensions especially important.

Seki, Bagchi, and Bagchi considered a hypercubic lattice with Gaussian-distributed site energies.  For a site \(i\) with energy \(U_i\), the equilibrium occupation probability is
\begin{equation}
\rho_i^{\rm eq}
=
\frac{\exp[-\beta U_i]}
{\left\langle \exp[-\beta U]\right\rangle},
\label{eq:SBB_eq_occ}
\end{equation}
where the angular brackets denote the average over the distribution of site energies.  The transition rates satisfy detailed balance.  It is then useful to introduce a symmetrized transition rate,
\begin{equation}
\Gamma_{ij}^{\rm sym}
=
\rho_i^{\rm eq}\Gamma_{ij}.
\label{eq:SBB_sym_rate}
\end{equation}
This symmetrization maps the transport problem onto an effective random-conductance problem, for which an effective-medium approximation can be constructed.

The central result of the Seki--Bagchi--Bagchi multidimensional treatment is the effective-medium self-consistency condition
\begin{equation}
\left\langle
\frac{
\Gamma_{\rm eff}-\Gamma_{\rm sym}
}{
(d-1)\Gamma_{\rm eff}+\Gamma_{\rm sym}
}
\right\rangle
=
0.
\label{eq:SBB_EMA}
\end{equation}
Here \(\Gamma_{\rm eff}\) is the effective transition rate in the medium, and the average is taken over the distribution of symmetrized rates.  This equation is the most important multidimensional result of the SBB analysis.  It should be distinguished from later approximate closed-form expressions, which are useful for interpretation but are not exact.

Equation~\eqref{eq:SBB_EMA} also makes clear why one dimension is special.  The denominator contains the factor \(d-1\).  Thus the structure of the effective-medium equation changes qualitatively when one goes from \(d=1\) to \(d>1\).  In one dimension, transport is controlled by bottlenecks in series.  In higher dimensions, parallel pathways become available.  This is the mathematical counterpart of the physical bypass argument.

Seki, Bagchi, and Bagchi obtained approximate analytical expressions from the effective-medium equation.  One such expression for the effective rate has the form
\begin{equation}
\frac{\Gamma_{\rm eff}}{\Gamma_0}
\simeq
\exp\left[
-\frac{\sigma^2}{2(k_{\rm B}T)^2}
-
\left(
\sqrt{\frac{2}{d}}
-
\frac{1}{\sqrt{2}}
\right)
\frac{\sqrt{\pi}\sigma}{k_{\rm B}T}
\right],
\label{eq:SBB_approx}
\end{equation}
where \(\sigma^2\) is the variance of the Gaussian site-energy distribution.  This expression captures the strong dimensionality dependence predicted by the effective-medium theory.  However, it should be regarded as an approximate analytical representation of the effective-medium result, not as an exact formula.

\subsection{Large-dimensional limit and the annealed hopping result}

The approximate expression in Eq.~(14) should not be used as the route to the infinite-dimensional limit. In the Seki--Bagchi--Bagchi treatment, the $(d\to\infty)$ result is obtained separately from the effective-medium self-consistency equation itself. Physically, this limit corresponds to the loss of memory of the previous jump. In one dimension, if a particle fails to cross a high barrier, it must return to the site it previously occupied. Thus successive jumps are strongly correlated. In high dimensions, by contrast, the number of possible neighboring sites becomes very large, and the probability of returning along the same unfavorable path becomes small. The transition events therefore become effectively independent in the infinite-dimensional limit.

\begin{equation}
\frac{\Gamma_{\rm eff}^{(\infty)}}{\Gamma_0}
=
\operatorname{erfc}
\left(
\frac{\sigma}{2k_BT}
\right).
\end{equation}

This result is already quoted as Eq.~(31) in the SBB paper. It is not obtained by extrapolating the approximate finite-dimensional expression, Eq.~(14), to $(d\to\infty)$. Rather, it follows directly from the $(d\to\infty)$ limit of the effective-medium self-consistency equation. SBB also showed this convergence explicitly: their Fig.~3 compares the finite-dimensional EMA solutions with the $(d=\infty)$ limiting curve and shows that the results approach this limiting expression as the dimensionality increases.

This distinction is important. The large-dimensional limit belongs to the discrete site-energy model with Miller--Abrahams hopping rates. It is therefore different in origin from Zwanzig's continuous rough-potential result. Zwanzig's expression follows from local smoothing of a continuous rugged potential, whereas the SBB $(d=\infty)$ expression follows from the loss of jump-to-jump memory in the high-dimensional limit of the discrete hopping model.

The main physical conclusion of Seki, Bagchi, and Bagchi is that the change from one to two dimensions is anomalously large.  The diffusion coefficient can increase substantially when the walker is allowed even one additional spatial direction, because bypass pathways become available.  The changes from two to three dimensions, and from three to higher dimensions, are comparatively weaker.  Thus, one-dimensional rugged-landscape diffusion is pathological in a precise sense: it overemphasizes serial bottlenecks and rare local traps that are partially avoided in higher-dimensional landscapes.

For the present work, the SBB analysis plays an important but limited role. The theory does not construct the full time-dependent propagator.  It also does not explicitly retain the dependence on the initial distribution of the particle, nor does it analyze higher moments of the displacement distribution such as those entering the non-Gaussian parameter.  These limitations motivate the propagator-based formulation developed in the following sections.
%
% ==============    SECTION 4   ===========================
%
%
\section{Propagator, Initial Preparation, and Experimental Observability}

The objective of this section is to develop a formulation that goes beyond the long-time diffusion coefficient. The Zwanzig, BBSB, and SBB treatments discussed above were primarily concerned with $D_{\rm eff}$, the asymptotic transport coefficient obtained after the system has reached its long-time diffusive regime. Zwanzig obtained such a coefficient by locally averaging roughness in a one-dimensional continuous Smoluchowski problem, BBSB obtained exact one-dimensional mean-first-passage-time results for a discrete Gaussian site-energy landscape, and SBB extended the calculation of the effective diffusion coefficient to higher dimensions using effective-medium theory. These results are central to the theory of rugged-landscape diffusion, but they do not describe the full finite-time displacement distribution. In particular, they do not retain the memory of how the particle was initially prepared, nor do they directly address the higher moments that characterize non-Gaussian diffusion.

We therefore adopt the propagator, or Green's-function, formulation. This approach keeps the initial distribution as part of the problem and allows one to define finite-time observables such as the apparent diffusion coefficient, the self-dynamic structure factor, and the non-Gaussian parameter. The long-time Gaussian form of the central part of the propagator is the natural expectation from central-limit and homogenization arguments once the particle has sampled many statistically distinct regions of the landscape. However, in a quenched rugged landscape this approach to the Gaussian limit can be slow and dimension dependent, because the effective steps are not all equivalent: some regions act as long-lived traps, and repeated visits to such regions generate memory. This issue is especially severe in one dimension, where barriers and traps occur in series and cannot be bypassed. In higher dimensions, alternate pathways make the central-limit picture more robust, although finite-time non-Gaussianity can still be large. The purpose of the formulation below is therefore twofold: first, to define the exact propagator-level quantities for a specified hopping model and initial preparation; and second, to analyze how the system approaches the long-time Gaussian diffusive regime through the relaxation of mobility-induced non-Gaussian fluctuations.

We consider a \(d\)-dimensional hypercubic lattice with lattice spacing \(b\).  The position of site \(i\) is denoted by \({\bf r}_i\), and the site energy is \(U_i\equiv U({\bf r}_i)\).  For a fixed realization of the quenched landscape, the probability \(P_i(t)\) of occupying site \(i\) evolves according to the master equation
\begin{equation}
\frac{dP_i(t)}{dt}
=
\sum_{j\in nn(i)}
\left[
\Gamma_{ji}P_j(t)-\Gamma_{ij}P_i(t)
\right],
\label{eq:master_section4}
\end{equation}
where \(nn(i)\) denotes the nearest neighbors of site \(i\).  The transition rates satisfy detailed balance with respect to the site energies.  A convenient choice, consistent with the BBSB and SBB lattice formulations, is the Miller--Abrahams rate
\begin{equation}
\Gamma_{ij}
=
\begin{cases}
\Gamma_0, & U_j\le U_i,\\
\Gamma_0\exp[-\beta(U_j-U_i)], & U_j>U_i.
\end{cases}
\label{eq:MA_section4}
\end{equation}

The disorder in the present theory is site-energy disorder.  The hopping rates are not independent random bond variables; they are determined by the neighboring site energies \(U_i\) and \(U_j\).

For a given landscape realization \(U\), Eq.~\eqref{eq:master_section4} may be written in matrix form as
\begin{equation}
\frac{d{\bf P}(t)}{dt}
=
{\bf W}[U]{\bf P}(t),
\label{eq:matrix_master_section4}
\end{equation}
where \({\bf W}[U]\) is the transition-rate matrix.  Its off-diagonal elements are
\begin{equation}
W_{ij}[U]
=
\Gamma_{ji},
\qquad i\ne j,
\label{eq:W_offdiag_section4}
\end{equation}
and its diagonal elements are
\begin{equation}
W_{ii}[U]
=
-\sum_{j\in nn(i)}\Gamma_{ij}.
\label{eq:W_diag_section4}
\end{equation}
The formal solution is
\begin{equation}
{\bf P}(t)
=
\exp\left[{\bf W}[U]t\right]{\bf P}(0).
\label{eq:formal_solution_section4}
\end{equation}

Thus the conditional propagator for a fixed landscape is
\begin{equation}
P(i,t|j,0;U)
=
\left[
\exp\left({\bf W}[U]t\right)
\right]_{ij}.
\label{eq:conditional_propagator_section4}
\end{equation}

If the initial distribution over sites is \(\rho_0(j|U)\), then the probability of occupying site \(i\) at time \(t\) is given by

\begin{equation}
P_i(t|\rho_0,U)
=
\sum_j
P(i,t|j,0;U)\rho_0(j|U).
\label{eq:Pi_rho0_section4}
\end{equation}

This is the first place at which the initial preparation enters explicitly.  In the earlier long-time effective-diffusion calculations, this dependence was either absent, hidden, or irrelevant to the final asymptotic coefficient.  Here it is retained from the beginning.

The displacement propagator, or the self part of the van Hove function, is defined as
\begin{equation}
G_s({\bf r},t|\rho_0,U)
=
\sum_{i,j}
\delta_{{\bf r},{\bf r}_i-{\bf r}_j}
P(i,t|j,0;U)\rho_0(j|U).
\label{eq:Gs_fixed_section4}
\end{equation}

The disorder-averaged propagator is then
\begin{equation}
G_s({\bf r},t|\rho_0)
=
\overline{
G_s({\bf r},t|\rho_0,U)
},
\label{eq:Gs_average_section4}
\end{equation}
where the overbar denotes an average over the ensemble of quenched rugged landscapes.

The site energies are taken to be drawn from a Gaussian random field with zero mean and spatial correlation function
\begin{equation}
\left\langle U({\bf r})U({\bf r}')\right\rangle
=
\sigma^2
\exp\left[
-\frac{|{\bf r}-{\bf r}'|^2}{2\xi^2}
\right].
\label{eq:Gaussian_field_section4}
\end{equation}

Here ${\bf r}$ denotes the displacement from the initially occupied site, not an absolute final position.  The initial site is averaged over the preparation distribution $\rho_0$, so all moments obtained from $G_s({\bf r},t|\rho_0)$ are preparation-averaged displacement moments.

On the lattice this becomes
\begin{equation}
\left\langle U_iU_j\right\rangle
=
\sigma^2
\exp\left[
-\frac{|{\bf r}_i-{\bf r}_j|^2}{2\xi^2}
\right].
\label{eq:lattice_Gaussian_field_section4}
\end{equation}
The parameters \(d\), \(\sigma\), and \(\xi\) specify the static ensemble of rugged landscapes.  The quantity \(\sigma\), or equivalently \(\epsilon\) in the notation used earlier, measures the energetic ruggedness.

%============== Discussion on beta_0/ Section 4  ==========

The initial distribution is an important physical ingredient.  We use the one-parameter family
\begin{equation}
\rho_0(i|U)
=
\frac{\exp[-\beta_0 U_i]}
{\sum_j \exp[-\beta_0 U_j]}.
\label{eq:rho0_section4}
\end{equation}
The parameter \(\beta_0\) characterizes the preparation protocol.  It should not be confused with the ruggedness of the landscape.  The ruggedness is fixed by the variance of the site-energy distribution, denoted here by \(\sigma^2\) or equivalently by \(\epsilon^2\).  The parameter \(\beta_0\) instead specifies how the initial ensemble samples that landscape.

Several limiting cases are useful.  The case \(\beta_0=0\) corresponds to uniform initial placement over the lattice.  The case \(\beta_0=\beta\) corresponds to equilibrium preparation at the bath temperature.  The case \(\beta_0>\beta\) describes an initially trap-biased ensemble, in which the particle is preferentially placed in low-energy regions.  The case \(\beta_0<0\) describes preparation biased toward high-energy regions.  This simple family allows us to study initial-state dependence, aging-like preparation effects, and the loss of memory of the initial basin.  In the following section we shall also use the dimensionless notation \(q_0=\beta_0\epsilon\), while \(q=\beta\epsilon\) measures the thermodynamic ruggedness.

%=====================================

%The parameter \(\beta_0\) characterizes the %preparation protocol.  The case \(\beta_0=0\) %corresponds to uniform initial placement over the %lattice.  The case \(\beta_0=\beta\) corresponds to %equilibrium preparation at the bath temperature.  %The case \(\beta_0>\beta\) describes an initially %trap-biased ensemble, in which the particle is %preferentially placed in low-energy regions.  The %case \(\beta_0<0\) describes preparation biased %toward high-energy regions.  This simple family %allows us to study initial-state dependence, aging-%like preparation effects, and the loss of memory of %the initial basin.

The mean displacement and mean-square displacement are obtained directly from the propagator:
\begin{equation}
\left\langle \Delta{\bf r}(t)\right\rangle_{\rho_0}
=
\sum_{\bf r}
{\bf r}\,
G_s({\bf r},t|\rho_0),
\label{eq:mean_disp_section4}
\end{equation}
and
\begin{equation}
\left\langle \Delta r^2(t)\right\rangle_{\rho_0}
=
\sum_{\bf r}
r^2
G_s({\bf r},t|\rho_0).
\label{eq:MSD_section4}
\end{equation}
The time-dependent apparent diffusion coefficient is defined as
\begin{equation}
D_{\rm app}(t;\rho_0)
=
\frac{1}{2d}
\frac{d}{dt}
\left[
\left\langle \Delta r^2(t)\right\rangle_{\rho_0}
-
\left|
\left\langle \Delta{\bf r}(t)\right\rangle_{\rho_0}
\right|^2
\right].
\label{eq:Dapp_section4}
\end{equation}

For an unbiased statistically homogeneous landscape, the disorder-averaged mean displacement vanishes, and Eq.~\eqref{eq:Dapp_section4} reduces to the time derivative of the mean-square displacement divided by \(2d\).

The conventional effective diffusion coefficient is recovered only if the long-time limit exists:
\begin{equation}
D_{\rm eff}
=
\lim_{t\rightarrow\infty}
D_{\rm app}(t;\rho_0).
\label{eq:Deff_section4}
\end{equation}

If the system is ergodic on the relevant time scale, this limiting value should become independent of \(\rho_0\).  However, in a rugged quenched landscape, especially at large ruggedness, small spatial correlation length, or low dimensionality, the approach to this limit can be very slow. \textit{ Thus the observed diffusion may remain preparation dependent over experimentally relevant times.}

The existence of the ergodic limit is an assumption about the long-time dynamics, not a consequence of the quenched model alone.

The experimental significance of Eq.~\eqref{eq:Dapp_section4} is clear.  A long-time diffusion coefficient can be extracted from the slope of the mean-square displacement in particle-tracking experiments, tracer diffusion, pulsed-field-gradient NMR, or simulations.  In ionic or charge-transport systems, the effective diffusion coefficient may be related to the conductivity through a Nernst--Einstein-type relation when the assumptions about charge carriers and correlations are satisfied.  Thus \(D_{\rm eff}/D_0\) provides a useful measure of the overall reduction of mobility by ruggedness.  However, \(D_{\rm eff}\) alone does not reveal whether transport is Gaussian or heterogeneous.

To access the shape of the displacement distribution, one must go beyond the mean-square displacement.  The Fourier transform of the propagator is the self-intermediate scattering function
\begin{equation}
F_s({\bf k},t|\rho_0)
=
\sum_{\bf r}
e^{i{\bf k}\cdot{\bf r}}
G_s({\bf r},t|\rho_0).
\label{eq:Fs_section4}
\end{equation}
This quantity is directly connected to scattering experiments.  In optical or colloidal systems, the propagator can often be obtained more directly from single-particle tracking or fluorescence microscopy, while in molecular systems the same information is commonly accessed through \(F_s({\bf k},t)\).  Thus both real-space and wave-vector-space measurements can probe the propagator.

For an isotropic displacement distribution, the small-\(k\) expansion is
\begin{equation}
F_s({\bf k},t)
=
1
-
\frac{k^2}{2d}
\left\langle \Delta r^2(t)\right\rangle
+
\frac{k^4}{8d(d+2)}
\left\langle \Delta r^4(t)\right\rangle
+
O(k^6).
\label{eq:Fs_moment_section4}
\end{equation}
The terms above are dependent on the initial distribution, $\rho_0$.

Equivalently, the cumulant expansion is
\begin{equation}
\ln F_s({\bf k},t)
=
-
\frac{k^2}{2d}
\left\langle \Delta r^2(t)\right\rangle
+
\frac{k^4}{8d(d+2)}
\left[
\left\langle \Delta r^4(t)\right\rangle
-
\frac{d+2}{d}
\left\langle \Delta r^2(t)\right\rangle^2
\right]
+
O(k^6).
\label{eq:Fs_cumulant_section4}
\end{equation}

The fourth-order term vanishes for Gaussian diffusion.  Therefore measurements of the small-\(k\) dependence of \(F_s({\bf k},t)\), or direct measurements of the displacement distribution, can be used to determine deviations from Gaussian behavior.

The standard non-Gaussian parameter in \(d\) dimensions is
\begin{equation}
\alpha_2(t|\rho_0)
=
\frac{d}{d+2}
\frac{
\left\langle \Delta r^4(t)\right\rangle_{\rho_0}
}{
\left\langle \Delta r^2(t)\right\rangle_{\rho_0}^2
}
-
1.
\label{eq:alpha2_section4}
\end{equation}

For a Gaussian propagator in \(d\) dimensions,
\begin{equation}
\left\langle \Delta r^4(t)\right\rangle
=
\frac{d+2}{d}
\left\langle \Delta r^2(t)\right\rangle^2,
\label{eq:Gaussian_fourth_section4}
\end{equation}
and hence \(\alpha_2(t)=0\).  A nonzero value of \(\alpha_2(t)\) measures the fourth-order deviation from Gaussian diffusion.  It is sensitive to intermittent trapping, dynamic heterogeneity, and incomplete sampling of the rugged landscape.  This makes \(\alpha_2(t)\) a more sensitive measure of transient ruggedness than \(D_{\rm eff}\) alone.

\subsection{Experimental and simulation studies of the non-Gaussian parameter}

The non-Gaussian parameter is not only a theoretical diagnostic. It is directly accessible whenever the displacement distribution, or equivalently the self part of the van Hove function, can be measured. Single-particle tracking experiments provide the most direct route because the full distribution of displacements can be constructed from trajectories. Scattering experiments provide a complementary route through the self-intermediate scattering function, $F_s(k,t)$, whose small-$k$ expansion contains both the mean-square displacement and the fourth-order cumulant. Thus $\alpha_2(t)$ probes information that is invisible in the long-time diffusion coefficient alone.

This point has become especially clear in recent studies of Brownian-yet-non-Gaussian and Fickian-yet-non-Gaussian diffusion. In these systems the mean-square displacement may grow linearly in time, while the displacement distribution remains markedly non-Gaussian because particles sample different local mobilities or because the same particle experiences a slowly fluctuating mobility environment \cite{Chechkin2017,Pastore2021,Miotto2021,Rusciano2022,PachecoPozo2023,Xu2024}. Related behavior has been observed or discussed in glass-forming liquids, colloidal systems, biological soft matter, interfacial diffusion, and protein sliding along DNA \cite{BagchiBlaineyXie2008,Blainey2009,Rusciano2022,Cherstvy2019}. These examples show that a theory based only on $D_{\rm eff}$ misses experimentally measurable finite-time information.

The present theory makes a specific prediction for such measurements. If ruggedness generates a spatially heterogeneous local mobility field, then the decay of $\alpha_2(t)$ should reflect the loss of correlation in the mobility sampled by the tagged particle. The predicted long-time forms,
$\alpha_2(t)\sim t^{-1/2}$ in one dimension, $\alpha_2(t)\sim(\ln t)/t$ in two dimensions, and $\alpha_2(t)\sim t^{-1}$ for $d>2$, can therefore be tested by measuring the displacement distribution over a broad time window. The one-dimensional case is especially relevant to sliding or curvilinear motion along polymers or DNA, where bypass pathways are absent or strongly restricted.

Simulations provide an even more direct test. For a prescribed quenched Gaussian energy landscape, one can generate trajectories using detailed-balance hopping rates and compute $G_s({\bf r},t)$, $D_{\rm app}(t)$, $F_s(k,t)$, $\alpha_2(t)$, and the local-mobility correlation function $C_D(t)$ from the same data. Simulations can also vary the dimensionality $d$, ruggedness amplitude, spatial correlation length $\xi$, and initial preparation $\rho_0$ independently. They therefore offer a clean way to test whether the observed non-Gaussian relaxation follows from site-induced mobility heterogeneity, as predicted here, rather than from hydrodynamic interactions, dynamic disorder, or other microscopic mechanisms.

The question then becomes how to relate \(\alpha_2(t)\) to the landscape.  We begin by defining the local escape rate
\begin{equation}
k_i[U]
=
\sum_{j\in nn(i)}
\Gamma_{ij}[U_i,U_j].
\label{eq:ki_section4}
\end{equation}
The associated site-induced local diffusivity is
\begin{equation}
D_i[U]
=
\frac{b^2}{2d}k_i[U].
\label{eq:Di_section4}
\end{equation}
Again, \(D_i[U]\) is not an independently assigned mobility.  It is a derived site-local quantity generated by the Gaussian site-energy landscape and the Miller--Abrahams hopping rule.

Along a stochastic trajectory \(X_t\), the particle samples the local diffusivity field
\begin{equation}
D_{\rm loc}(t)
=
D_{X_t}[U].
\label{eq:Dloc_section4}
\end{equation}
Thus, instead of treating diffusion as governed by a single number, we regard the particle as sampling a spatially heterogeneous mobility field induced by the site-energy disorder.

For a fixed landscape and initial distribution, the autocorrelation of the sampled local mobility is
\begin{equation}
C_D(t|\rho_0,U)
=
\sum_{i,j}
\delta D_i\,
P(i,t|j,0;U)\,
\rho_0(j|U)\,
\delta D_j,
\label{eq:CD_fixed_section4}
\end{equation}
where
\begin{equation}
\delta D_i
=
D_i[U]-\left\langle D\right\rangle_{\rho_0,U},
\label{eq:deltaD_section4}
\end{equation}
and
\begin{equation}
\left\langle D\right\rangle_{\rho_0,U}
=
\sum_i
\rho_0(i|U)D_i[U].
\label{eq:Dmean_fixed_section4}
\end{equation}
The disorder-averaged mobility autocorrelation is
\begin{equation}
C_D(t|\rho_0)
=
\overline{
C_D(t|\rho_0,U)
}.
\label{eq:CD_average_section4}
\end{equation}
This expression is formally exact.  It says that mobility memory is the correlation between the local diffusivity sampled initially and the local diffusivity sampled at time \(t\), averaged over both trajectories and rugged landscapes.

For equilibrium preparation, the process is stationary for each fixed landscape.  Then
\begin{equation}
C_D(t|U)
=
\sum_{i,j}
\delta D_i\,
P(i,t|j,0;U)\,
\rho_j^{\rm eq}\,
\delta D_j.
\label{eq:CD_eq_section4}
\end{equation}
Because detailed balance holds, the generator can be symmetrized, and the equilibrium mobility correlation has the spectral representation
\begin{equation}
C_D(t|U)
=
\sum_{n\ge 1}
A_n(U)\exp[-\lambda_n(U)t].
\label{eq:CD_spectral_section4}
\end{equation}
The zero mode is absent because the correlation is written in terms of fluctuations \(\delta D_i\).  Slow relaxation modes associated with traps and bottlenecks therefore appear directly as slowly decaying contributions to \(C_D(t)\).

The connection to \(\alpha_2(t)\) can be made through the time-integrated local mobility
\begin{equation}
{\cal A}(t)
=
\int_0^t ds\,D_{\rm loc}(s).
\label{eq:A_section4}
\end{equation}
In a heterogeneous-diffusivity approximation, the non-Gaussian parameter is approximately the relative variance of this integrated mobility:
\begin{equation}
\alpha_2(t)
\simeq
\frac{
{\rm Var}[{\cal A}(t)]
}{
\left\langle {\cal A}(t)\right\rangle^2
}.
\label{eq:alpha_A_section4}
\end{equation}
For equilibrium preparation, or more generally when \(D_{\rm loc}(t)\) may be treated as stationary over the time interval of interest,
\begin{equation}
{\rm Var}[{\cal A}(t)]
=
2\int_0^t ds\,(t-s)C_D(s).
\label{eq:VarA_section4}
\end{equation}
Therefore,
\begin{equation}
\alpha_2(t)
\simeq
\frac{
2
}{
t^2\left\langle D\right\rangle^2
}
\int_0^t ds\,(t-s)C_D(s).
\label{eq:alpha_CD_section4}
\end{equation}

This equation is a central result of the present formulation.  It relates the fourth-order non-Gaussianity of the displacement distribution to the time correlation of the local mobility sampled by the particle.  It also shows why \(D_{\rm eff}\) alone is incomplete: the mean diffusivity controls the second moment, but the fluctuations and memory of the mobility field control \(\alpha_2(t)\).

In the long-time coarse-grained form used below, the average sampled diffusivity $\langle D\rangle$ may be identified with $D_{\rm eff}$; we retain the notation $\langle D\rangle$ in Eq.~(50) to emphasize its origin as an average over the local diffusivity field.

Equation~\eqref{eq:alpha_CD_section4} is also experimentally useful.  The left-hand side, \(\alpha_2(t)\), can be obtained from the measured displacement distribution or from the fourth-order term in the small-\(k\) expansion of \(F_s({\bf k},t)\).  The right-hand side expresses this observable in terms of a mobility-memory kernel.  Thus the theory predicts that non-Gaussianity is large when the particle remains correlated with its initial mobility environment for a long time.  Trap-biased initial preparation should enhance this effect, while high-dimensional bypass pathways and spatial smoothing should reduce it.

To proceed analytically, we introduce the spatial covariance of the site-induced local diffusivity field,
\begin{equation}
C_D^{\rm sp}({\bf r})
=
\overline{
\delta D({\bf r};U)\delta D({\bf 0};U)
}.
\label{eq:CDsp_section4}
\end{equation}
This covariance is induced by the correlated site-energy field.  It is not an independently imposed mobility or bond-disorder correlation.

In a weak-disorder or annealed propagator approximation, the time-dependent mobility correlation sampled by the particle may be estimated as a convolution of this spatial covariance with a coarse-grained propagator:
\begin{equation}
C_D(t)
\simeq
\int d{\bf r}\,
C_D^{\rm sp}({\bf r})
G_0({\bf r},t).
\label{eq:CD_convolution_section4}
\end{equation}
Here
\begin{equation}
G_0({\bf r},t)
=
\frac{1}{(4\pi D_{\rm eff}t)^{d/2}}
\exp\left[
-\frac{r^2}{4D_{\rm eff}t}
\right]
\label{eq:G0_section4}
\end{equation}
is the coarse-grained Gaussian propagator.  The approximation in Eq.~\eqref{eq:CD_convolution_section4} neglects detailed quenched trapping correlations, but it captures the physically important idea that the particle loses memory of its initial mobility environment by diffusing away from it.

If the spatial mobility covariance is approximately Gaussian,
\begin{equation}
C_D^{\rm sp}({\bf r})
=
C_D(0)
\exp\left[
-\frac{r^2}{2\xi_D^2}
\right],
\label{eq:CDsp_Gaussian_section4}
\end{equation}
then Eq.~\eqref{eq:CD_convolution_section4} gives
\begin{equation}
C_D(t)
\simeq
C_D(0)
\left[
1+\frac{2D_{\rm eff}t}{\xi_D^2}
\right]^{-d/2}.
\label{eq:CD_closed_section4}
\end{equation}
Here \(\xi_D\) is the correlation length of the site-induced mobility field.  It is related to, but not necessarily identical with, the energy correlation length \(\xi\), because the local diffusivity depends on energy differences and escape rates rather than on the site energy alone.

Combining Eq.~\eqref{eq:alpha_CD_section4} with Eq.~\eqref{eq:CD_closed_section4} gives
\begin{equation}
\alpha_2(t)
\simeq
\frac{
2C_D(0)
}{
t^2\left\langle D\right\rangle^2
}
\int_0^t ds\,
(t-s)
\left[
1+\frac{2D_{\rm eff}s}{\xi_D^2}
\right]^{-d/2}.
\label{eq:alpha_main_section4}
\end{equation}
This is the principal analytical result of the present section.  It expresses the non-Gaussian parameter in terms of the amplitude of mobility heterogeneity, the mobility correlation length, the effective long-time diffusion coefficient, and the dimensionality.  The result predicts that non-Gaussianity is controlled not only by how much the average diffusion coefficient is reduced, but also by how spatially correlated and persistent the induced mobility field is.

Several physical conclusions follow immediately.  First, the amplitude \(C_D(0)/\langle D\rangle^2\) measures the strength of local mobility heterogeneity.  This amplitude should increase with energetic ruggedness.  Second, the time scale \(\xi_D^2/(2D_{\rm eff})\) is the time required for the particle to diffuse across a correlated mobility domain.  Third, the factor \(d/2\) gives an explicit dimensionality dependence: mobility memory decays more slowly in one dimension than in two or three dimensions.  Thus Eq.~\eqref{eq:alpha_main_section4} gives an analytical expression of the bypass idea discussed in connection with the SBB work.

The initial distribution enters this picture through both \(C_D(0)\) and the early-time part of \(C_D(t)\).  A trap-biased initial condition, corresponding to \(\beta_0>\beta\), overweights low-mobility regions and should increase the magnitude and persistence of non-Gaussianity.  A high-energy initial preparation, corresponding to \(\beta_0<0\), should produce a different early-time response, with faster initial motion followed by relaxation into the typical rugged landscape.  Thus the theory predicts that \(\alpha_2(t)\), not only \(D_{\rm eff}\), should depend on the preparation protocol at intermediate times.

The propagator formulation also allows us to quantify memory of the initial preparation directly.  One possible measure is the distance between the propagator generated from \(\rho_0\) and that generated from equilibrium preparation:
\begin{equation}
{\cal M}_{\rho_0}(t)
=
\sum_{\bf r}
\left|
G_s({\bf r},t|\rho_0)
-
G_s({\bf r},t|\rho_{\rm eq})
\right|.
\label{eq:memory_measure_section4}
\end{equation}
If \({\cal M}_{\rho_0}(t)\) decays rapidly to zero, the system quickly forgets its initial preparation.  If it decays slowly, the system is formally ergodic but practically preparation dependent on accessible time scales.  If it fails to decay, the system retains genuine long-time memory of the initial condition.

The experimental consequences are therefore broader than a change in the average diffusion coefficient.  Conductivity or long-time tracer diffusion can measure the overall departure of \(D_{\rm eff}\) from \(D_0\).  Optical and single-particle measurements can provide the displacement distribution and hence \(\alpha_2(t)\).  Scattering experiments can access the self-intermediate scattering function and its cumulants.  Simulations can evaluate the full propagator, the site-induced mobility field, and the preparation-dependent memory function directly.  The present formulation connects all of these observables by showing how ruggedness, correlations, dimensionality, and preparation enter the propagator.

The present work focuses on quenched rugged landscapes with spatial correlations.  The same framework can be extended to dynamic disorder by allowing the landscape itself to fluctuate in time.  In that case the energy field would be characterized by a spatiotemporal correlation function of the form
\begin{equation}
\left\langle U({\bf r},t)U({\bf r}',s)\right\rangle
=
\epsilon^2
C_s(|{\bf r}-{\bf r}'|)
C_t(|t-s|),
\label{eq:spatiotemporal_section4}
\end{equation}
where \(C_s\) describes spatial correlations and \(C_t\) describes temporal correlations.  The associated propagator would then require an additional average over histories of the fluctuating landscape.  This dynamic-disorder extension is not developed here, but the propagator formulation makes clear how it should enter.

In summary, the propagator-based formulation replaces the older focus on a single asymptotic number by a hierarchy of time-dependent observables:
\begin{equation}
G_s({\bf r},t|\rho_0),
\qquad
F_s({\bf k},t|\rho_0),
\qquad
D_{\rm app}(t;\rho_0),
\qquad
\alpha_2(t;\rho_0),
\qquad
C_D(t|\rho_0).
\label{eq:observable_hierarchy_section4}
\end{equation}
These observables retain information about dimensionality, ruggedness, spatial correlations, and initial preparation.  The older effective diffusion constant is recovered only as the final long-time limit, when such a limit exists.  The central question is therefore not only how \(D_{\rm eff}\) depends on \(d\), \(\sigma\), and \(\xi\), but how the full propagator crosses over from a preparation-dependent, trap-dominated, non-Gaussian regime to an asymptotic diffusive regime.
%
%
%%  ==================  SECTION 5   ===================
%
%
\section{From Site-Energy Ruggedness to Non-Gaussian Diffusion}

\subsection{analytical Expressions}

The preceding section formulated the diffusion problem at the level of the propagator and showed that the non-Gaussian parameter is controlled by a mobility-memory correlation.  In particular, the approximate relation
\begin{equation}
\alpha_2(t)
\simeq
\frac{
2C_D(0)
}{
t^2\langle D\rangle^2
}
\int_0^t ds\,
(t-s)
\left[
1+\frac{2D_{\rm eff}s}{\xi_D^2}
\right]^{-d/2}
\label{eq:alpha_section5_start}
\end{equation}
expresses \(\alpha_2(t)\) in terms of the amplitude of local mobility heterogeneity, \(C_D(0)\), the mobility correlation length \(\xi_D\), the long-time diffusion coefficient \(D_{\rm eff}\), and the dimensionality \(d\).  The objective of the present section is to estimate the quantities entering Eq.~\eqref{eq:alpha_section5_start} in terms of the microscopic ruggedness of the site-energy landscape.

The disorder remains strictly site disorder.  The random variables are the site energies \(U_i\), which are Gaussian and spatially correlated.  The hopping rates are not independent random bonds.  They are deterministic functions of the neighboring site energies through the Miller--Abrahams rule.  Thus the sequence of ideas is
\begin{equation}
\hbox{site-energy disorder}
\quad \Longrightarrow \quad
\hbox{site-dependent escape rates}
\quad \Longrightarrow \quad
\hbox{site-induced mobility field}.
\label{eq:site_to_mobility_section5}
\end{equation}
This distinction is important because the present theory remains in the same class of site-energy models considered in BBSB and SBB.

We use \(\epsilon\) to denote the width of the Gaussian site-energy distribution.  This is the ruggedness parameter in the sense of Zwanzig and BBSB.  In the preceding section the same variance was denoted by \(\sigma^2\); here \(\epsilon^2\) and \(\sigma^2\) should be understood as the same site-energy variance.  We introduce the dimensionless ruggedness
\begin{equation}
q=\beta\epsilon,
\label{eq:q_section5}
\end{equation}
and the nearest-neighbor correlation parameter
\begin{equation}
c
=
\exp\left[
-\frac{b^2}{2\xi^2}
\right].
\label{eq:c_section5}
\end{equation}
Thus \(c=0\) is the uncorrelated nearest-neighbor limit, while \(c\rightarrow 1\) is the strongly correlated, locally smooth limit.

The site energies satisfy
\begin{equation}
\left\langle U_i\right\rangle=0,
\qquad
\left\langle U_i^2\right\rangle=\epsilon^2,
\label{eq:site_variance_section5}
\end{equation}
and, for nearest-neighbor sites \(i\) and \(j\),
\begin{equation}
\left\langle U_iU_j\right\rangle=\epsilon^2 c.
\label{eq:site_correlation_section5}
\end{equation}
The Miller--Abrahams rate is
\begin{equation}
\Gamma_{ij}
=
\begin{cases}
\Gamma_0, & U_j\le U_i,\\
\Gamma_0\exp[-\beta(U_j-U_i)], & U_j>U_i.
\end{cases}
\label{eq:MA_rate_section5}
\end{equation}
All averages in this section are therefore Gaussian averages over site energies.

\subsection{The limit of equilibrium preparation and local rate averages}

We first consider equilibrium preparation.  This is the natural reference case because the stochastic process is stationary for each fixed rugged landscape.  The initial occupation is then proportional to \(\exp[-\beta U_i]\), and the relevant nearest-neighbor energy difference is
\begin{equation}
\Delta U_{ij}=U_j-U_i.
\label{eq:DeltaU_section5}
\end{equation}
Under equilibrium weighting of the initial site, \(\Delta U_{ij}\) remains Gaussian.  Its mean is
\begin{equation}
\left\langle \Delta U_{ij}\right\rangle_{\rm eq}
=
\beta\epsilon^2(1-c),
\label{eq:DeltaU_mean_eq_section5}
\end{equation}
and its variance is
\begin{equation}
\left\langle
\left(
\Delta U_{ij}
-
\left\langle \Delta U_{ij}\right\rangle_{\rm eq}
\right)^2
\right\rangle
=
2\epsilon^2(1-c).
\label{eq:DeltaU_var_eq_section5}
\end{equation}
Equations~\eqref{eq:DeltaU_mean_eq_section5} and \eqref{eq:DeltaU_var_eq_section5} show how spatial correlation enters the local hopping problem.  The variance of the nearest-neighbor energy increment is reduced by the factor \(1-c\).  Thus correlations smooth the local roughness increments.

Let
\begin{equation}
\Phi(x)
=
\frac{1}{\sqrt{2\pi}}
\int_{-\infty}^{x}dy\,e^{-y^2/2}
\label{eq:Phi_section5}
\end{equation}
be the normal cumulative distribution function.  The equilibrium average of the Miller--Abrahams rate is
\begin{equation}
\frac{\left\langle \Gamma\right\rangle_{\rm eq}}{\Gamma_0}
=
\operatorname{erfc}
\left[
\frac{q}{2}\sqrt{1-c}
\right].
\label{eq:Gamma_mean_eq_section5}
\end{equation}
This is the first useful analytical result.  It gives the local equilibrium-averaged hopping rate in terms of the ruggedness \(q=\beta\epsilon\) and the nearest-neighbor correlation \(c\).

Several limits are immediate.  In the uncorrelated limit \(c=0\),
\begin{equation}
\frac{\left\langle \Gamma\right\rangle_{\rm eq}}{\Gamma_0}
=
\operatorname{erfc}\left(\frac{q}{2}\right).
\label{eq:Gamma_eq_uncorr_section5}
\end{equation}
In the strongly correlated limit \(c\rightarrow 1\),
\begin{equation}
\frac{\left\langle \Gamma\right\rangle_{\rm eq}}{\Gamma_0}
\rightarrow 1.
\label{eq:Gamma_eq_corr_section5}
\end{equation}
Thus local spatial smoothness removes the nearest-neighbor hopping penalty.  This is the local-mobility analogue of the correlated BBSB result, where spatial correlations suppress sharp trapping configurations and restore Zwanzig-like behavior.

The initial apparent diffusion coefficient follows from the short-time motion.  Since a nearest-neighbor hop has length \(b\),
\begin{equation}
D_{\rm app}(0^+)
=
b^2\left\langle \Gamma\right\rangle_{\rm eq}.
\label{eq:Dapp_initial_section5}
\end{equation}
If
\begin{equation}
D_0=b^2\Gamma_0,
\label{eq:D0_lattice_section5}
\end{equation}
then
\begin{equation}
\frac{D_{\rm app}(0^+)}{D_0}
=
\frac{\left\langle \Gamma\right\rangle_{\rm eq}}{\Gamma_0}
=
\operatorname{erfc}
\left[
\frac{q}{2}\sqrt{1-c}
\right].
\label{eq:Dapp_ratio_section5}
\end{equation}
This is not the long-time diffusion coefficient.  It is the initial local mobility sampled by an equilibrium ensemble.  Its value lies in showing explicitly how ruggedness and spatial correlation enter the early-time propagator.

\subsection{Mobility dispersion due to ruggedness}

The non-Gaussian parameter depends not only on the mean mobility but also on mobility fluctuations.  We therefore need the second moment of the local Miller--Abrahams rate.  The same two-site Gaussian average gives
\begin{equation}
\frac{\left\langle \Gamma^2\right\rangle_{\rm eq}}{\Gamma_0^2}
=
\Phi\left[
-q\sqrt{\frac{1-c}{2}}
\right]
+
\exp\left[
2q^2(1-c)
\right]
\Phi\left[
-3q\sqrt{\frac{1-c}{2}}
\right].
\label{eq:Gamma_second_eq_section5}
\end{equation}
This expression is again a site-disorder result.  It is obtained by averaging over the correlated Gaussian pair \((U_i,U_j)\), not by assigning independent random values to the link \(ij\).

The local escape rate from site \(i\) is
\begin{equation}
k_i[U]
=
\sum_{j\in nn(i)}
\Gamma_{ij}[U_i,U_j],
\label{eq:ki_section5}
\end{equation}
and the site-induced local diffusivity is
\begin{equation}
D_i[U]
=
\frac{b^2}{2d}k_i[U].
\label{eq:Di_section5}
\end{equation}
The zero-time mobility variance is
\begin{equation}
C_D(0)
=
\overline{
\left[
D_i[U]-\overline{D_i}
\right]^2
}.
\label{eq:CD0_def_section5}
\end{equation}
Because \(D_i[U]\) is a sum over \(2d\) outgoing rates, the exact variance contains correlations among different exits from the same site.  These correlations arise naturally in a site-disorder model because all outgoing rates share the same central site energy \(U_i\).

If \(z=2d\) is the coordination number, the exact structure is
\begin{equation}
C_D(0)
=
\frac{b^4}{z^2}
\left[
z
\left(
\left\langle \Gamma^2\right\rangle_{\rm eq}
-
\left\langle \Gamma\right\rangle_{\rm eq}^2
\right)
+
z(z-1)
\left(
\left\langle \Gamma_{ij}\Gamma_{ik}\right\rangle_{\rm eq}^{(j\ne k)}
-
\left\langle \Gamma\right\rangle_{\rm eq}^2
\right)
\right].
\label{eq:CD0_exact_structure_section5}
\end{equation}
The second term involves a three-site Gaussian average over \((U_i,U_j,U_k)\).  It is retained here to make clear that no independent bond-disorder assumption is being made.

A useful first approximation is to neglect the cross-correlation among different exits from the same site after the local equilibrium weighting has been included:
\begin{equation}
\left\langle \Gamma_{ij}\Gamma_{ik}\right\rangle_{\rm eq}^{(j\ne k)}
\simeq
\left\langle \Gamma\right\rangle_{\rm eq}^2.
\label{eq:exit_decoupling_section5}
\end{equation}
This is a local-exit decoupling approximation, not a change of model.  With this approximation,
\begin{equation}
C_D(0)
\simeq
\frac{b^4}{2d}
\left[
\left\langle \Gamma^2\right\rangle_{\rm eq}
-
\left\langle \Gamma\right\rangle_{\rm eq}^2
\right].
\label{eq:CD0_approx_section5}
\end{equation}
Since
\begin{equation}
\left\langle D\right\rangle
=
b^2\left\langle \Gamma\right\rangle_{\rm eq},
\label{eq:Dmean_section5}
\end{equation}
we obtain
\begin{equation}
\frac{C_D(0)}{\left\langle D\right\rangle^2}
\simeq
\frac{1}{2d}
\left[
\frac{
\left\langle \Gamma^2\right\rangle_{\rm eq}
}{
\left\langle \Gamma\right\rangle_{\rm eq}^2
}
-1
\right].
\label{eq:CD0_ratio_section5}
\end{equation}
Equations~\eqref{eq:Gamma_mean_eq_section5}, \eqref{eq:Gamma_second_eq_section5}, and \eqref{eq:CD0_ratio_section5} provide an explicit analytical estimate of the amplitude of site-induced mobility heterogeneity in terms of ruggedness and spatial correlation.

\subsection{Equilibrium result for the non-Gaussian parameter}

We now substitute Eq.~\eqref{eq:CD0_ratio_section5} into the mobility-memory result.  This gives
\begin{equation}
\alpha_2(t)
\simeq
\frac{1}{d}
\left[
\frac{
\left\langle \Gamma^2\right\rangle_{\rm eq}
}{
\left\langle \Gamma\right\rangle_{\rm eq}^2
}
-1
\right]
\frac{1}{t^2}
\int_0^t ds\,
(t-s)
\left[
1+\frac{2D_{\rm eff}s}{\xi_D^2}
\right]^{-d/2}.
\label{eq:alpha_eq_final_section5}
\end{equation}
This is the central analytical result of the equilibrium part of the theory.  It expresses the non-Gaussian parameter in terms of the microscopic ruggedness through the first two moments of the Miller--Abrahams rate.  The ruggedness \(q\) and the site-energy correlation \(c\) enter through Eqs.~\eqref{eq:Gamma_mean_eq_section5} and \eqref{eq:Gamma_second_eq_section5}.  The dimensionality enters in two ways: through the prefactor \(1/d\), which reflects averaging over more escape directions, and through the memory-decay factor with exponent \(d/2\).

It is useful to define the mobility-memory time
\begin{equation}
\tau_D
=
\frac{\xi_D^2}{2D_{\rm eff}}.
\label{eq:tauD_section5}
\end{equation}
For times short compared with \(\tau_D\), the particle has not moved far enough to lose memory of its initial mobility environment.  Then
\begin{equation}
\left[
1+\frac{s}{\tau_D}
\right]^{-d/2}
\simeq 1,
\label{eq:short_memory_factor_section5}
\end{equation}
and Eq.~\eqref{eq:alpha_eq_final_section5} gives the short-time plateau
\begin{equation}
\alpha_2(t)
\simeq
\frac{1}{2d}
\left[
\frac{
\left\langle \Gamma^2\right\rangle_{\rm eq}
}{
\left\langle \Gamma\right\rangle_{\rm eq}^2
}
-1
\right],
\qquad
t\ll \tau_D.
\label{eq:alpha_short_section5}
\end{equation}
This limit shows that the initial non-Gaussianity is controlled by the relative variance of the local mobility field.  A perfectly homogeneous landscape gives no local mobility variance and therefore no non-Gaussianity from this mechanism.

At long times, the memory kernel in Eq.~\eqref{eq:alpha_eq_final_section5} decays as
\begin{equation}
C_D(t)\sim t^{-d/2}.
\label{eq:CD_long_section5}
\end{equation}
Thus the decay of \(\alpha_2(t)\) depends on dimensionality.  In one dimension,
\begin{equation}
\alpha_2(t)\sim t^{-1/2},
\qquad d=1.
\label{eq:alpha_long_d1_section5}
\end{equation}
In two dimensions,
\begin{equation}
\alpha_2(t)\sim \frac{\ln t}{t},
\qquad d=2.
\label{eq:alpha_long_d2_section5}
\end{equation}
For dimensions larger than two,
\begin{equation}
\alpha_2(t)\sim t^{-1},
\qquad d>2.
\label{eq:alpha_long_dgt2_section5}
\end{equation}
These limiting forms provide a simple interpretation of dimensionality.  Mobility memory decays slowly in one dimension because the walker repeatedly revisits the same local environment.  In higher dimensions, bypass pathways and spatial spreading accelerate the loss of mobility memory.  This is the propagator-level counterpart of the SBB conclusion that one-dimensional rugged-landscape diffusion is anomalous.

Equation~\eqref{eq:alpha_eq_final_section5} also clarifies the role of spatial correlations.  Increasing \(\xi\) increases the nearest-neighbor correlation \(c\), thereby reducing the effective local energy increment.  This reduces the variance of the local rates and hence lowers the amplitude of \(\alpha_2(t)\).  On the other hand, the mobility correlation length \(\xi_D\) controls the time scale over which mobility memory decays.  Thus spatial correlations can reduce the amplitude of local trapping while also extending the spatial scale over which mobility fluctuations are correlated.  This competition should be visible in numerical simulations.
\subsection{Study of nonequilibrium initial preparation}

The equilibrium analysis above gives the cleanest analytical baseline.  However, one of the motivations for the propagator approach is that it allows us to retain explicit initial-state dependence.  Zwanzig's original rough-potential calculation, the BBSB mean-first-passage-time treatment, and the SBB effective-medium theory were all primarily concerned with long-time effective diffusion constants.  They did not formulate the full propagator with an independently variable initial distribution.  The present approach therefore permits a new question: how does the early and intermediate time propagator depend on preparation?

We introduce
\begin{equation}
q_0=\beta_0\epsilon,
\label{eq:q0_section5}
\end{equation}
where \(q_0\) is a preparation parameter, not a second ruggedness.  The ruggedness of the landscape is still \(q=\beta\epsilon\).  The case \(q_0=0\) corresponds to uniform initial placement, \(q_0=q\) corresponds to equilibrium preparation, \(q_0>q\) corresponds to trap-biased preparation, and \(q_0<0\) corresponds to high-energy-biased preparation.

For a statistically homogeneous large system, the average over the initially occupied site can be written as
\begin{equation}
\left\langle A\right\rangle_{\beta_0}
=
\frac{
\left\langle e^{-\beta_0 U_i}A\right\rangle
}{
\left\langle e^{-\beta_0 U_i}\right\rangle
}.
\label{eq:biased_average_section5}
\end{equation}
Under this biased average, the nearest-neighbor energy difference remains Gaussian, with mean
\begin{equation}
\left\langle \Delta U_{ij}\right\rangle_{\beta_0}
=
\beta_0\epsilon^2(1-c),
\label{eq:DeltaU_mean_beta0_section5}
\end{equation}
and variance
\begin{equation}
\left\langle
\left(
\Delta U_{ij}
-
\left\langle \Delta U_{ij}\right\rangle_{\beta_0}
\right)^2
\right\rangle
=
2\epsilon^2(1-c).
\label{eq:DeltaU_var_beta0_section5}
\end{equation}
The preparation changes the mean energy step sampled initially, but not the variance of the underlying Gaussian energy increment.

The preparation-dependent mean Miller--Abrahams rate is
\begin{equation}
\frac{\left\langle \Gamma\right\rangle_{\beta_0}}{\Gamma_0}
=
\Phi\left[
-q_0\sqrt{\frac{1-c}{2}}
\right]
+
\exp\left[
(q^2-qq_0)(1-c)
\right]
\Phi\left[
(q_0-2q)\sqrt{\frac{1-c}{2}}
\right].
\label{eq:Gamma_mean_beta0_section5}
\end{equation}
This reduces to Eq.~\eqref{eq:Gamma_mean_eq_section5} when \(q_0=q\).  For uniform initial preparation, \(q_0=0\), one obtains
\begin{equation}
\frac{\left\langle \Gamma\right\rangle_{\rm unif}}{\Gamma_0}
=
\frac{1}{2}
+
\exp\left[
q^2(1-c)
\right]
\Phi\left[
-2q\sqrt{\frac{1-c}{2}}
\right].
\label{eq:Gamma_uniform_section5}
\end{equation}
The initial apparent diffusivity is therefore preparation dependent:
\begin{equation}
\frac{D_{\rm app}(0^+;\beta_0)}{D_0}
=
\frac{\left\langle \Gamma\right\rangle_{\beta_0}}{\Gamma_0}.
\label{eq:Dapp_beta0_section5}
\end{equation}

The second moment is
\begin{equation}
\frac{\left\langle \Gamma^2\right\rangle_{\beta_0}}{\Gamma_0^2}
=
\Phi\left[
-q_0\sqrt{\frac{1-c}{2}}
\right]
+
\exp\left[
(4q^2-2qq_0)(1-c)
\right]
\Phi\left[
(q_0-4q)\sqrt{\frac{1-c}{2}}
\right].
\label{eq:Gamma_second_beta0_section5}
\end{equation}
In the same local-exit decoupling approximation, this gives
\begin{equation}
\frac{C_D(0;\beta_0)}{\left\langle D\right\rangle_{\beta_0}^2}
\simeq
\frac{1}{2d}
\left[
\frac{
\left\langle \Gamma^2\right\rangle_{\beta_0}
}{
\left\langle \Gamma\right\rangle_{\beta_0}^2
}
-1
\right].
\label{eq:CD0_beta0_section5}
\end{equation}

For \(q_0\ne q\), the process is not stationary at the initial time.  Therefore the exact relation between \(\alpha_2(t)\) and mobility fluctuations should involve a two-time mobility correlation,
\begin{equation}
C_D(s,u;\beta_0)
=
\left\langle
\delta D_{\rm loc}(s)\delta D_{\rm loc}(u)
\right\rangle_{\beta_0}.
\label{eq:CD_twotime_section5}
\end{equation}
The corresponding integrated-mobility approximation is
\begin{equation}
\alpha_2(t;\beta_0)
\simeq
\frac{
\int_0^t ds\int_0^t du\,
C_D(s,u;\beta_0)
}{
\left[
\int_0^t ds\,
\left\langle D_{\rm loc}(s)\right\rangle_{\beta_0}
\right]^2
}.
\label{eq:alpha_twotime_section5}
\end{equation}
This is the proper nonequilibrium generalization.  It shows explicitly why initial-state dependence is a more difficult problem than equilibrium mobility memory.

As a first analytical approximation for early and intermediate times, one may assume that nonequilibrium preparation mainly changes the initial amplitude of the mobility heterogeneity, while the subsequent loss of memory is still controlled by diffusive escape from mobility domains.  This gives
\begin{equation}
\alpha_2(t;\beta_0)
\simeq
\frac{1}{d}
\left[
\frac{
\left\langle \Gamma^2\right\rangle_{\beta_0}
}{
\left\langle \Gamma\right\rangle_{\beta_0}^2
}
-1
\right]
\frac{1}{t^2}
\int_0^t ds\,
(t-s)
\left[
1+\frac{2D_{\rm eff}s}{\xi_D^2}
\right]^{-d/2}.
\label{eq:alpha_beta0_final_section5}
\end{equation}
Equation~\eqref{eq:alpha_beta0_final_section5} is not an exact nonequilibrium identity.  It is a controlled phenomenological extension of the equilibrium result, useful because it displays explicitly how the measured non-Gaussian parameter can depend on initial preparation.

This preparation dependence is potentially observable.  In simulations, \(q_0\) can be imposed directly.  In experiments, different initial subensembles may be accessed by optical selection, spatial localization, trapping, or selective excitation.  Uniform initial placement, equilibrium preparation, trap-biased preparation, and high-energy-biased preparation should give different early-time values of \(D_{\rm app}(t)\) and \(\alpha_2(t)\), even if the long-time \(D_{\rm eff}\) eventually becomes independent of the initial state.

\subsection{Connection to numerical work}

The analytical results of this section are approximate.  Equations~\eqref{eq:Gamma_mean_eq_section5} and \eqref{eq:Gamma_second_eq_section5} are two-site Gaussian averages and are therefore local.  Equation~\eqref{eq:CD0_ratio_section5} uses a local-exit decoupling approximation.  Equation~\eqref{eq:alpha_eq_final_section5} uses a Gaussian approximation for the spatial covariance of the site-induced mobility field and a coarse-grained diffusive propagator to describe loss of mobility memory.

These approximations should be tested directly by numerical propagation of the master equation on correlated Gaussian site-energy landscapes.  The numerical work should compute the full propagator \(G_s({\bf r},t|\rho_0)\), the apparent diffusion coefficient \(D_{\rm app}(t)\), the non-Gaussian parameter \(\alpha_2(t)\), and the mobility correlation \(C_D(t)\).  It should then compare the measured \(\alpha_2(t)\) with Eq.~\eqref{eq:alpha_eq_final_section5} for equilibrium preparation and with Eq.~\eqref{eq:alpha_beta0_final_section5} for nonequilibrium preparation.  This provides the natural transition to the numerical analysis in the next section.
%
%====================  Section 6  ============================

\section{Explicit Analytical Results in One Dimension}

The general theory developed above predicts a general d-dimensional hierarchy in the relaxation of the non-Gaussian parameter. Before carrying out full numerical simulations in two and three dimensions, it is useful to examine the one-dimensional case analytically. This case is special because the long-time diffusion coefficient is available from the BBSB result, and because the absence of bypass pathways makes one-dimensional rugged-landscape diffusion maximally sensitive to local mobility memory. The calculation below therefore serves as a benchmark for the propagator-level theory.

\subsection{Why the one-dimensional case is special}

We first examine the one-dimensional system separately. This case is useful for two reasons. First, the one-dimensional rugged landscape is known to be special because a particle cannot bypass a deep trap or a large local barrier. This is the physical origin of the anomalous behavior emphasized in the BBSB and SBB analyses. Second, in one dimension the long-time diffusion coefficient is already known analytically to high accuracy from the BBSB mean-first-passage-time treatment. 

Thus the one-dimensional case provides a clean baseline for testing the propagator-level theory of non-Gaussianity.  We use the Miller--Abrahams nearest-neighbor rate

\begin{equation}
\Gamma_{ij}
=
\begin{cases}
\Gamma_0, & U_j\le U_i,\\
\Gamma_0\exp[-\beta(U_j-U_i)], & U_j>U_i.
\end{cases}
\label{eq:MA_rate_1d}
\end{equation}

We choose the lattice spacing (b) and the bare nearest-neighbor rate $(\Gamma_0)$ as the microscopic length and time scales. With the convention that $(\Gamma_0)$ is the hopping rate to each nearest neighbor, the bare diffusion coefficient is

\begin{equation}
D_0=b^2\Gamma_0.
\label{eq}
\end{equation}

We therefore use the dimensionless time

\begin{equation}
\tau=\frac{D_0t}{b^2}.
\label{eq}
\end{equation}
Equivalently, if (b=1) and $(\Gamma_0=1)$, then $(D_0=1)$ and $(\tau=t)$.

We also set $(k_{\rm B}T=1)$ in the numerical illustrations, so that the dimensionless ruggedness is

\begin{equation}
q=\beta\epsilon=\epsilon.
\label{eq}
\end{equation}

Thus, for the uncorrelated one-dimensional model, the only nontrivial control parameter is the ruggedness amplitude $(\epsilon)$.

\subsection{BBSB expression for the effective diffusion coefficient}

For an uncorrelated one-dimensional Gaussian site-energy landscape, the BBSB mean-first-passage-time result for the effective diffusion coefficient is

\begin{equation}
D_{eff} = D_{\rm BBSB}
=
D_0 \exp(-\beta^2\epsilon^2)
\left[
1+\operatorname{erf}\left(\frac{\beta\epsilon}{2}\right)
\right]^{-1}.
\label{eq:BBSB_Deff}
\end{equation}

by the additional error-function factor. The correction arises from the discrete one-dimensional hopping geometry. In one dimension, a particle trapped in a deep minimum bounded by higher-energy neighbors has no alternate path. Such rare three-site traps strongly influence the long-time mean-first-passage time and lead to the BBSB correction.

Eq. (110) will be used here as the one-dimensional value of $(D_{\rm eff})$. This avoids the need for trajectory simulations in the first part of the numerical analysis and allows us to focus directly on the prediction for $(\alpha_2(t))$.

\subsection{Local rate moments and mobility heterogeneity}

We define the normal cumulative distribution function as
\begin{equation}
\Phi(x)
=
\frac{1}{\sqrt{2\pi}}
\int_{-\infty}^{x} dy\,
\exp\left(-\frac{y^2}{2}\right).
\label{eq:normal_cdf}
\end{equation}

The function $\Phi(x)$ is the normal cumulative distribution function.  We also define the relative rate dispersion
\begin{equation}
R(\epsilon)
=
\frac{\langle \Gamma^2\rangle_{\rm eq}}
{\langle \Gamma\rangle_{\rm eq}^2}
-1 .
\label{eq:rate_dispersion_R}
\end{equation}

The short-time limiting value of the non-Gaussian parameter is then
\begin{equation}
\alpha_2(0^+)
=
\frac{1}{2}R(\epsilon).
\label{eq:alpha2_short_time}
\end{equation}

Thus the initial amplitude of non-Gaussianity is directly controlled by the relative variance of the local Miller--Abrahams rate.

\subsection{Analytical expression for ${(\alpha_2(t))}$ in one dimension}

We now specialize to the mobility-memory expression in one dimension. For the first analytical estimate we take the mobility-correlation length to be the lattice spacing,

\begin{equation}
\xi_D=b.
\label{eq}
\end{equation}

This is the natural starting choice for the uncorrelated site-energy landscape.  More refined estimates of $\xi_D$ can be extracted later from numerical simulations of the mobility field.  With this choice, the dimensionless mobility-memory parameter is

\begin{equation}
a(\epsilon)
=
2\frac{D_{\rm eff}^{1D}}{D_0}.
\label{eq:mobility_memory_parameter}
\end{equation}

With the natural one-dimensional choice $\xi_D=b$, the memory-decay rate is set by diffusion across one mobility-correlation length, giving $a(\epsilon)=2D_{\rm eff}^{1D}/D_0$.
The appearance of $(D_{\rm eff}^{1D})$ in $(a(\epsilon))$ reflects the fact that the particle loses memory of its initial mobility environment by diffusing away from it.

The time integral can be evaluated analytically as
\begin{equation}
I(\tau;\epsilon)
=
\int_0^\tau du\,(\tau-u)\,[1+a(\epsilon)u]^{-1/2}
=
\frac{4}{3a^2}
\left[
(1+a\tau)^{3/2}-1
\right]
-\frac{2\tau}{a}.
\label{eq:alpha2_memory_integral_1d}
\end{equation}
Equations~\eqref{eq:rate_dispersion_R}--\eqref{eq:alpha2_memory_integral_1d} give a closed analytical approximation for the one-dimensional non-Gaussian parameter.  No trajectory simulation is required at this stage.

\subsection{Limiting behavior and crossover}

The short-time limit follows immediately from Eq. 116. For $(\tau\ll a^{-1})$, $\left[1+a u \right]^{-1/2} \simeq 1, $
and hence,

\begin{equation}
I(\tau;\epsilon)
\simeq
\frac{\tau^2}{2}.
\label{eq}
\end{equation}

Therefore,

\begin{equation}
\alpha_2(\tau)
\simeq
\frac{1}{2}R(\epsilon),
\qquad
\tau\ll a^{-1}.
\label{eq}
\end{equation}

The early-time plateau is therefore determined by the relative variance of the local rate distribution.

At long times, $(a\tau\gg1)$, Eq.~\eqref{eq} gives

\begin{equation}
I(\tau;\epsilon)
\simeq
\frac{4}{3\sqrt{a}}
\tau^{3/2}.
\label{eq}
\end{equation}

Thus we have,

\begin{equation}
\alpha_2(\tau)
\simeq
\frac{4R(\epsilon)}
{3\sqrt{a(\epsilon)}}
\tau^{-1/2}.
\label{eq}
\end{equation}

We now use $(a=2D_{\rm eff}^{1D}/D_0)$, this may also be written as
\begin{equation}
\alpha_2(\tau)
\simeq
\frac{4R(\epsilon)}
{3\sqrt{2D_{\rm eff}^{1D}/D_0}}
\tau^{-1/2}.
\label{eq}
\end{equation}

Note that the dimensionless time $\tau$ is defined by $\tau=\frac{D_{0} t}{b^2}$.

The slow $t^{-1/2}$ decay is the one-dimensional mobility-memory anomaly. It is the propagator-level counterpart of the anomalous one-dimensional behavior already known from the BBSB and SBB analyses of $D_{\rm eff}$. In one dimension, the walker repeatedly revisits the same local environment, and the memory of local mobility fluctuations is lost only slowly.

The crossover time marks the time at which the non-Gaussian parameter  $\alpha_{2}(\tau)$ leaves its initial plateau and begins to approach the asymptotic one-dimensional power-law decay.

The crossover time is thus set by

\begin{equation}
\tau_D
=
\frac{1}{a(\epsilon)}
=
\frac{1}{2D_{\rm eff}^{1D}/D_0}.
\label{eq:tauD_1d}
\end{equation}
Since $D_{\rm eff}^{1D}/D_0$ decreases rapidly with $\epsilon$, the crossover time increases rapidly with ruggedness. This explains why strongly rugged systems can remain close to their initial non-Gaussian plateau over very long observation times.

\subsection{Numerical illustration of the analytical one-dimensional result}

Figure 1 shows the one-dimensional analytical prediction for $(\alpha_2(\tau))$ for following four ruggedness amplitudes:

\begin{equation}
\epsilon=0.2,;0.5,;1.0,;2.0.
\label{eq}
\end{equation}
The larger value $(\epsilon=4.0)$ is not included in this figure because the corresponding $(D_{\rm eff})$ is extremely small and the curve remains close to its plateau over the plotted time window.

Panel A shows the raw $(\alpha_2(\tau))$ on ordinary axes. The increase of the initial plateau with ruggedness is evident. 
Panel B shows the same data on log--log axes and includes a reference slope $-1/2$, corresponding to the long-time one-dimensional prediction in Eqs. 120-121.
Panel C shows the normalized quantity $(\alpha_2(\tau)/\alpha_2(0^+))$, which removes the amplitude difference and emphasizes the relaxation of mobility memory.

%==================================== Figure 1 three panels ===============================
\begin{figure}[t]
    \centering
    \includegraphics[width=\textwidth]{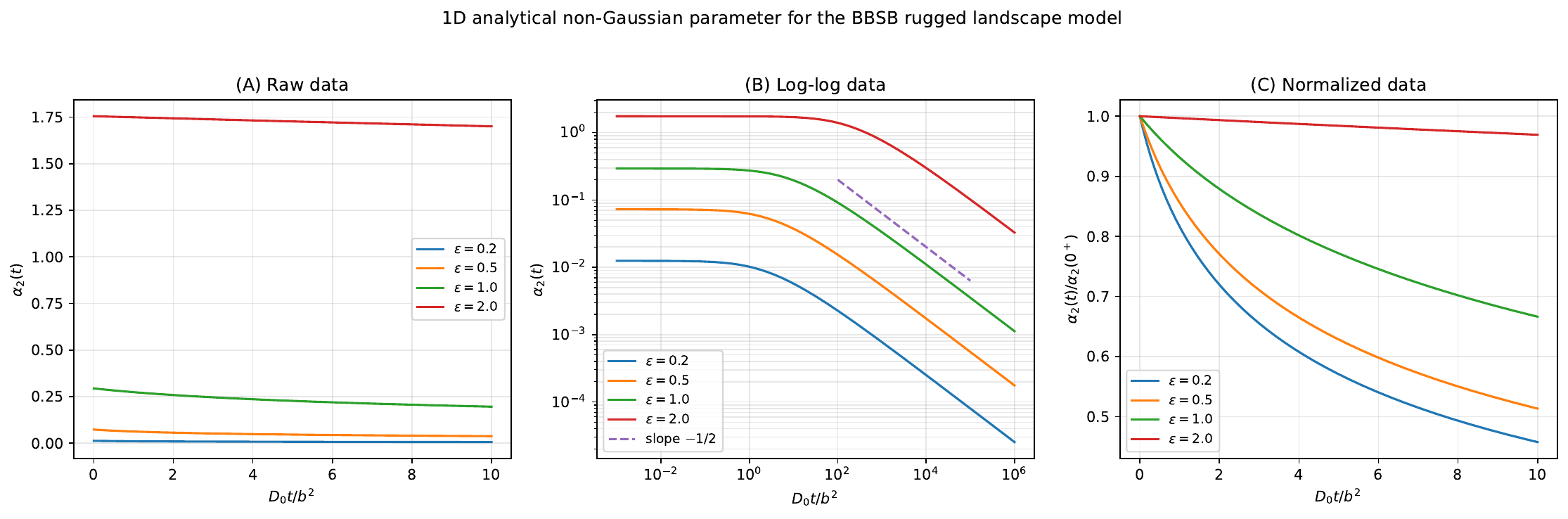}
    \caption{
    One-dimensional analytical prediction for the non-Gaussian parameter
    $\alpha_2(t)$ in the uncorrelated BBSB rugged landscape model.\cite{BBSB2014}
    The dimensionless time is $D_0t/b^2$, and the ruggedness values are
    $\epsilon=0.2,0.5,1.0,2.0$.
    (A) Raw $\alpha_2(t)$ on ordinary axes.
    (B) Raw $\alpha_2(t)$ on log--log axes, showing comparison with the
    asymptotic one-dimensional $t^{-1/2}$ decay.
    (C) Normalized non-Gaussian parameter
    $\alpha_2(t)/\alpha_2(0^+)$, emphasizing the relaxation of mobility memory.
    }
    \label{fig:alpha2_1D_three_panel}
\end{figure}

%===============================  End Figure 1 ==================================

The three panels emphasize different aspects of the same result. The raw ordinary plot shows the magnitude of non-Gaussianity. The log--log plot displays the slow asymptotic decay. The normalized plot shows how the relaxation shape changes with ruggedness after the initial amplitude is scaled out. Together, they show that ruggedness affects both the size of the non-Gaussian parameter and the time scale over which it decays.

\subsection{Implications for higher-dimensional numerical work}

The one-dimensional result is special because $(D_{\rm eff}^{1D}/D_0)$ is available analytically from the BBSB expression. In higher dimensions, the SBB effective-medium theory gives valuable guidance for $(D_{\rm eff})$, but the full propagator and the non-Gaussian parameter are most naturally obtained by numerical propagation or trajectory simulation. The one-dimensional calculation therefore serves as a benchmark.

In the next step, we shall extend the calculation to two and three dimensions. The central question will be whether the non-Gaussian parameter shows the predicted dimensional hierarchy: slow $(t^{-1/2})$ decay in one dimension, marginal $((\ln t)/t)$ behavior in two dimensions, and faster $(t^{-1})$ decay in higher dimensions. This will allow us to test directly whether the anomalous dimensionality dependence known for $(D_{\rm eff})$ also appears in the full propagator and its fourth-order cumulant.

\section{Spatiotemporal Mobility Correlations and the Self-Dynamic Structure Factor}

The preceding sections introduced the local mobility field and its time correlation along the tagged-particle trajectory.  It is useful to state explicitly the underlying spatiotemporal correlation function from which this memory arises.  Let
\begin{equation}
\delta D({\bf r},t)
=
D({\bf r},t)-\langle D\rangle
\end{equation}
denote the fluctuation of the local diffusivity.  The full spatiotemporal mobility correlation function is then
\begin{equation}
C_D({\bf r},t)
=
\left\langle
\delta D({\bf 0},0)\delta D({\bf r},t)
\right\rangle .
\end{equation}
For a strictly quenched landscape, $D({\bf r},t)$ is time independent in the laboratory frame, but the tagged particle samples different values of $D({\bf r})$ as it moves through the landscape.  Thus the time dependence of the sampled mobility memory arises from the propagator of the tagged particle.  In a dynamic landscape, by contrast, $D({\bf r},t)$ also has an intrinsic temporal dependence.

The normalized mobility-memory kernel is defined as
\begin{equation}
M_D(t)
=
\frac{C_D(t)}{C_D(0)} .
\end{equation}
In the one-dimensional reduced form, with $\xi_D=b$ and $\tau=D_0t/b^2$, this kernel becomes
\begin{equation}
M_D(\tau)
=
[1+a(\epsilon)\tau]^{-1/2},
\end{equation}
where
\begin{equation}
a(\epsilon)
=
2\frac{D_{\rm eff}^{1D}(\epsilon)}{D_0}.
\end{equation}
Thus $a(\epsilon)$ is not a separate microscopic memory function.  It is the dimensionless decay rate appearing in the one-dimensional form of the normalized mobility-memory kernel.

The same physics appears naturally in the self-dynamic structure factor.  The self part of the van Hove function is the displacement propagator $G_s({\bf r},t)$, and its Fourier transform is
\begin{equation}
F_s(k,t)
=
\left\langle
\exp[i{\bf k}\cdot\Delta{\bf r}(t)]
\right\rangle
=
\int d{\bf r}\,
\exp(i{\bf k}\cdot{\bf r})G_s({\bf r},t).
\end{equation}
For an isotropic system, the small-$k$ expansion is
\begin{equation}
F_s(k,t)
=
1
-
\frac{k^2}{2d}
\left\langle
\Delta r^2(t)
\right\rangle
+
\frac{k^4}{8d(d+2)}
\left\langle
\Delta r^4(t)
\right\rangle
+
O(k^6).
\end{equation}
Equivalently, the cumulant expansion gives
\begin{equation}
\ln F_s(k,t)
=
-
\frac{k^2}{2d}
\left\langle
\Delta r^2(t)
\right\rangle
+
\frac{k^4}{8d(d+2)}
\left[
\left\langle
\Delta r^4(t)
\right\rangle
-
\frac{d+2}{d}
\left\langle
\Delta r^2(t)
\right\rangle^2
\right]
+
O(k^6).
\end{equation}
The expression in square brackets is the fourth-order deviation from Gaussian diffusion.  Using the definition
\begin{equation}
\alpha_2(t)
=
\frac{d}{d+2}
\frac{\left\langle \Delta r^4(t)\right\rangle}
{\left\langle \Delta r^2(t)\right\rangle^2}
-
1,
\end{equation}
the cumulant expansion can be written as
\begin{equation}
\ln F_s(k,t)
=
-
\frac{k^2}{2d}
\left\langle
\Delta r^2(t)
\right\rangle
+
\frac{k^4}{8d^2}
\alpha_2(t)
\left\langle
\Delta r^2(t)
\right\rangle^2
+
O(k^6).
\end{equation}
This relation shows that $\alpha_2(t)$ is directly the coefficient of the leading non-Gaussian correction to $\ln F_s(k,t)$.

For ordinary Gaussian diffusion, $F_s(k,t)=\exp(-D_{\rm eff}k^2t)$ and $\alpha_2(t)=0$.  In a rugged landscape, however, the mean-square displacement may already be close to its diffusive form while the fourth cumulant remains nonzero.  If $\langle \Delta r^2(t)\rangle\simeq 2dD_{\rm eff}t$, then
\begin{equation}
\ln F_s(k,t)
\simeq
-
D_{\rm eff}k^2t
+
\frac{1}{2}
\alpha_2(t)D_{\rm eff}^2k^4t^2
+
O(k^6).
\end{equation}
Thus measurements of $F_s(k,t)$ at small but finite $k$ provide a direct route to the non-Gaussian parameter.  The spatiotemporal mobility correlation controls the time dependence of $\alpha_2(t)$, while the self-dynamic structure factor provides the experimentally accessible representation of the same fourth-order physics.

This is the rest of the section only, with all displayed equations complete.

% =====================  7 Conclusion =================
%
\section {Conclusion}

Before we summarize and discuss the results, it is useful to state clearly the scope of the theory. The propagator $G_s({\bf r},t|\rho_0,U)$ for a specified quenched landscape and initial distribution is an exact object of the master-equation formulation. Its disorder average $G_s({\bf r},t|\rho_0)$, and the moments and cumulants derived from it, are also well-defined once the ensemble of landscapes is specified. Thus the propagator formulation itself does not rely on a Gaussian approximation.

The approximation enters at the next stage, where we interpret the spatially varying escape rates as a mobility field sampled by the moving particle. This mobility-memory construction is designed to describe how heterogeneous trapping produces finite-time non-Gaussianity and how this non-Gaussianity decays as the particle samples more of the landscape. In dimensions three and higher, this decay is expected to lead naturally to the usual Gaussian diffusive propagator, provided that a normal diffusion coefficient $D_{\rm eff}$ exists. This is the regime in which the asymptotic power laws derived here are most directly applicable.

The lower-dimensional cases require more care. In one dimension, traps and barriers occur in series and cannot be bypassed, so recurrence and the order of the limits $t\to\infty$ and $L\to\infty$ can strongly affect the approach to the Gaussian regime. Two dimensions is intermediate: bypass paths exist, but recurrence remains important and can produce marginal behavior. Thus the power laws derived here should be read as predictions for the homogenized diffusive regime: most robust for $d\ge 3$, marginal in $d=2$, and potentially fragile in $d=1$.

We next summarize the main results of this paper. We have presented a propagator-based theory of tagged-particle diffusion in a multidimensional rugged energy landscape with quenched spatial correlations. The main objective was to go beyond the traditional description in terms of only the long-time diffusion coefficient, $D_{\rm eff}$, and to analyze finite-time observables that retain information about dimensionality, local mobility fluctuations, and initial preparation.

The principal results are the following.

(i) We formulated the diffusion problem at the level of the self-propagator, $G_s({\bf r},t|\rho_0)$, for a specified initial distribution $\rho_0$. Here ${\bf r}$ denotes the displacement from the initially occupied site, while the initial site itself is averaged over the preparation distribution. This formulation makes explicit the dependence of finite-time diffusion on the way the tagged particle is initially placed in the rugged landscape. Uniform, equilibrium, trap-biased, and high-energy-biased preparations can therefore be treated within the same framework.

(ii) We emphasized that the self-dynamic structure factor, $F_s(k,t|\rho_0)$, is the experimentally accessible Fourier representation of the same propagator. Its small-$k$ expansion contains the second and fourth displacement moments, while the corresponding cumulant expansion brings in the non-Gaussian parameter $\alpha_2(t;\rho_0)$. Thus $\alpha_2(t;\rho_0)$ is not merely a formal fourth-moment diagnostic. It is the coefficient of the leading non-Gaussian correction to $\ln F_s(k,t|\rho_0)$. When the mean-square displacement has already reached its diffusive form, this correction appears as a term proportional to $\alpha_2(t)D_{\rm eff}^2k^4t^2$ in $\ln F_s(k,t)$. This provides a direct connection between the present propagator theory and scattering measurements.

(iii) We showed that a Gaussian site-energy landscape, together with detailed-balance hopping rates, generates a spatially heterogeneous local mobility field. Thus the relevant disorder is not an independently imposed bond disorder or mobility disorder. Instead, the mobility heterogeneity is induced by the underlying rugged site-energy landscape through the local escape rates. We derived analytical expressions for the mean and variance of these local hopping rates, and hence for the amplitude of the local mobility dispersion. These expressions show explicitly how energetic ruggedness and spatial correlation length control the strength of mobility heterogeneity.

(iv) We related the non-Gaussian parameter $\alpha_2(t)$ to the time correlation of the local diffusivity sampled by the tagged particle. In the local-diffusivity representation, the particle samples a time-dependent local diffusivity along its trajectory, and the time integral of this local diffusivity controls the displacement statistics. The non-Gaussian parameter is then governed by the relative variance of this integrated diffusivity. Equivalently, $\alpha_2(t)$ is controlled by the mobility-memory correlation function $C_D(t)$. This result provides a direct connection between microscopic ruggedness, mobility fluctuations, and the fourth-order cumulant of the displacement distribution. It also shows why $D_{\rm eff}$ alone is insufficient: the long-time diffusion coefficient measures the asymptotic second moment, whereas $\alpha_2(t)$ probes the persistence and decay of local mobility fluctuations.

(v) We introduced the spatial covariance of the site-induced mobility field and used it to estimate the time dependence of the sampled mobility memory. For a Gaussian mobility covariance with correlation length $\xi_D$, convolution with the coarse-grained propagator gives a normalized mobility-memory kernel $M_D(t)=C_D(t)/C_D(0)$ containing the characteristic factor $[1+2D_{\rm eff}t/\xi_D^2]^{-d/2}$. Here $C_D(t)$ is the mobility-memory correlation function, while $M_D(t)$ is its normalized kernel. This distinction is important because $C_D(t)$ contains both the amplitude and the memory of local mobility fluctuations, whereas $M_D(t)$ describes only the normalized decay of memory.

(vi) The most important result is that $\alpha_2(t)$ exhibits a dimension-dependent power-law relaxation. The mobility-memory kernel implies that $\alpha_2(t)\sim t^{-1/2}$ in one dimension, $\alpha_2(t)\sim(\ln t)/t$ in two dimensions, and $\alpha_2(t)\sim t^{-1}$ for $d>2$. This relaxation of the non-Gaussian parameter is not obtained from the effective diffusion coefficient alone. It is a propagator-level signature of rugged-landscape diffusion.

These results extend the earlier emphasis on the anomalous dimensionality dependence of $D_{\rm eff}$ to the full time-dependent propagator and its fourth-order cumulant. In one dimension, the tagged particle repeatedly revisits the same local environment and loses memory of its mobility surroundings only slowly. In higher dimensions, bypass pathways and spatial spreading accelerate the loss of this memory. Thus the special role of one dimension appears not only in the effective diffusion coefficient, but also in the relaxation of non-Gaussian fluctuations.

The theory also clarifies the relation to earlier work. Zwanzig's result concerns diffusion in a continuous rough potential after local smoothing of the roughness. The BBSB correction arises in a discrete one-dimensional site-energy landscape, where rare three-site traps produce a correction absent in Zwanzig's treatment. The Seki--Bagchi--Bagchi effective-medium theory showed that the long-time diffusion coefficient has a strong anomalous dimensionality dependence, especially in going from one to two dimensions, and that the high-dimensional limit approaches the corresponding $d=\infty$ result of the Miller--Abrahams hopping model. The present work builds on these results but shifts the focus from $D_{\rm eff}$ alone to the full propagator, the self-dynamic structure factor, and the non-Gaussian parameter.

The present theory suggests several direct tests. In simulations, one can generate quenched Gaussian site-energy landscapes with controlled ruggedness and correlation length, propagate the corresponding master equation or stochastic trajectories, and measure $G_s({\bf r},t|\rho_0)$, $F_s(k,t|\rho_0)$, $D_{\rm app}(t)$, $C_D(t)$, $M_D(t)$, and $\alpha_2(t)$ directly. The preparation dependence can be tested by comparing uniform, equilibrium, trap-biased, and high-energy-biased initial ensembles. Experimentally, the predicted relaxation of $\alpha_2(t)$ may be examined in systems where single-particle tracking, scattering, fluorescence measurements, or numerical trajectories provide access to the displacement distribution beyond the mean-square displacement. In scattering experiments, the most direct signature is the leading non-Gaussian correction to $\ln F_s(k,t)$ at small but finite $k$. An interesting case would be comparison of self-dynamic structure factor of stable glasses with those formed by rapid quenching. \cite {Ediger1, Kushal1, Kushal2, WolynesUltraStable}
The difference can reveal the extent of ruggedness, and also initial preparation dependence.

Several extensions will be worthwhile. The present treatment assumes quenched rugged landscapes with spatial correlations. Dynamically fluctuating landscapes would require a genuinely spatiotemporal correlation function for the energy or mobility field. It would also be useful to test the predicted power-law relaxation of $\alpha_2(t)$ in explicit two- and three-dimensional simulations, and to examine how the mobility-correlation length $\xi_D$ is related quantitatively to the energy-correlation length $\xi$. These extensions should help connect the present propagator-level theory more directly to heterogeneous diffusion in soft matter, biomolecular motion, disordered solids, and glassy systems.

%====================  Appendices   ================

\section* {Appendix A : Summary of Symbols,Their Physical Meanings and Technical Details}

\renewcommand{\theequation}{A\arabic{equation}}
\setcounter{equation}{0}

This appendix collects the main symbols used in the paper and states their physical meaning. The purpose is to distinguish clearly among energetic ruggedness, spatial correlations, mobility correlations, and the memory kernel that controls the decay of the non-Gaussian parameter.

\begin{enumerate}

\item $\epsilon$ denotes the Gaussian ruggedness amplitude of the energy landscape. It is the standard deviation of the rough part of the energy landscape. Thus $\epsilon^2$ is the variance of the Gaussian site-energy distribution. In Zwanzig's notation, this is the roughness strength that enters the result $D_Z=D_0\exp(-\beta^2\epsilon^2)$. In the present lattice formulation, the same symbol denotes the width of the Gaussian site-energy distribution.

\item $q=\beta\epsilon$ denotes the dimensionless ruggedness. It measures the energy roughness in units of thermal energy $k_BT$. Large $q$ corresponds to strong energetic trapping and slow diffusion.

\item $\xi$ denotes the spatial correlation length of the Gaussian site-energy landscape.  It characterizes how rapidly the site energies decorrelate in space.  For a Gaussian correlated landscape,
\begin{equation}
\langle U({\bf r})U({\bf 0})\rangle
=
\epsilon^2
\exp\left(-\frac{r^2}{2\xi^2}\right).
\end{equation}
Thus $\xi$ is an energy-correlation length.

\item $\xi_D$ denotes the correlation length of the induced local mobility field. It is not necessarily identical to $\xi$. The energy field $U({\bf r})$ generates site-dependent hopping rates, and these hopping rates generate a spatially heterogeneous local diffusivity field $D({\bf r})$. The length $\xi_D$ measures the spatial range over which this induced local mobility field remains correlated. In the simplest one-dimensional estimate used in Sec.~6, we take $\xi_D=b$, where $b$ is the lattice spacing.

\item $D_0$ denotes the bare diffusion coefficient in the absence of energetic ruggedness. On the one-dimensional lattice, with nearest-neighbor hopping rate $\Gamma_0$ and lattice spacing $b$, we use
\begin{equation}
D_0=b^2\Gamma_0 .
\end{equation}
It is the reference diffusion coefficient before the rugged energy landscape slows the motion.

\item $D_{\rm eff}$ denotes the long-time effective diffusion coefficient in the rugged landscape. It is obtained from the long-time limit of the mean-square displacement, or equivalently from the long-time limit of the apparent diffusion coefficient. It is the final coarse-grained transport coefficient after the particle has sampled the rugged landscape for a long time.

\item $D_{\rm eff}^{1D}(\epsilon)$ denotes the one-dimensional effective diffusion coefficient in the BBSB uncorrelated Gaussian site-energy landscape. It depends on the ruggedness $\epsilon$. In the one-dimensional analytical illustration, this quantity is used to set the time scale for loss of mobility memory.

\item $D_{\rm app}(t)$ denotes the apparent time-dependent diffusion coefficient, defined from the time derivative of the mean-square displacement. Unlike $D_{\rm eff}$, it retains finite-time information and can depend on the initial preparation.

\item $D_i$ or $D({\bf r})$ denotes the site-induced local diffusivity. It is not assigned independently. It is generated by the local escape rate from a site:
\begin{equation}
D_i=\frac{b^2}{2d}k_i ,
\end{equation}
where $k_i$ is the total escape rate from site $i$. Thus the local mobility field is derived from the Gaussian site-energy landscape and the hopping rule.

\item $C_D(t)$ denotes the time correlation function of the local diffusivity sampled by the moving particle. It is the central mobility-memory correlation function of the theory. In words, it measures how strongly the local diffusivity experienced by the particle at time $t$ remains correlated with the local diffusivity sampled initially. If $C_D(t)$ decays slowly, the particle retains memory of its initial mobility environment for a long time.

\item $C_D(0)$ denotes the equal-time variance of the local diffusivity field sampled by the particle.  The ratio $C_D(0)/\langle D\rangle^2$ measures the strength of local mobility heterogeneity.

\item $C_D^{\rm sp}(r)$ denotes the spatial covariance of the site-induced local diffusivity field.  It measures how local mobility fluctuations at two points separated by distance $r$ are correlated.  A simple Gaussian approximation is
\begin{equation}
C_D^{\rm sp}(r)
=
C_D(0)
\exp\left(-\frac{r^2}{2\xi_D^2}\right).
\end{equation}
Thus $C_D^{\rm sp}(r)$ contains both the amplitude of mobility fluctuations and the mobility-correlation length $\xi_D$.

\item $C_D(t;\epsilon,\xi_D)$ denotes the mobility-memory correlation function viewed as a function of ruggedness and mobility-correlation length. The ruggedness $\epsilon$ controls the amplitude of mobility fluctuations and the reduction of $D_{\rm eff}$, while $\xi_D$ controls the spatial range over which mobility remains correlated.

\item The memory kernel denotes the normalized time-dependent factor that describes the loss of mobility memory. In the Gaussian mobility-covariance approximation, diffusion away from the initial mobility environment gives
\begin{equation}
\frac{C_D(t)}{C_D(0)}
\simeq
\left[
1+\frac{2D_{\rm eff}t}{\xi_D^2}
\right]^{-d/2}.
\end{equation}
This factor is the memory kernel used in the theory. It is not an independently postulated function; it follows from convolving the spatial mobility covariance with the coarse-grained Gaussian propagator.

\item $a(\epsilon)$ denotes the dimensionless memory-decay rate used in the one-dimensional analytical illustration. It is not itself the memory function. It appears after the one-dimensional memory kernel is written in terms of the dimensionless time $\tau=D_0t/b^2$ and the simplifying choice $\xi_D=b$. Starting from
\begin{equation}
\left[
1+\frac{2D_{\rm eff}^{1D}t}{\xi_D^2}
\right]^{-1/2},
\end{equation}
and using $\xi_D=b$ and $\tau=D_0t/b^2$, one obtains
\begin{equation}
\left[
1+a(\epsilon)\tau
\right]^{-1/2},
\end{equation}
where
\begin{equation}
a(\epsilon)=2\frac{D_{\rm eff}^{1D}(\epsilon)}{D_0}.
\end{equation}
Thus $a(\epsilon)$ measures the rate at which the particle diffuses out of its initial mobility environment in dimensionless time. Since increasing ruggedness lowers $D_{\rm eff}^{1D}$, larger $\epsilon$ gives smaller $a(\epsilon)$ and therefore slower memory decay.

\item $\tau$ denotes the dimensionless time used in the one-dimensional analytical section:
\begin{equation}
\tau=\frac{D_0t}{b^2}.
\end{equation}
It measures time in units of the bare microscopic diffusion time over one lattice spacing.

\item $\tau_D$ denotes the dimensionless crossover time for loss of mobility memory in one dimension:
\begin{equation}
\tau_D=a(\epsilon)^{-1}.
\end{equation}
It marks the crossover from the initial plateau of the non-Gaussian parameter to the onset of the long-time asymptotic power-law decay. Since $a(\epsilon)$ decreases with increasing ruggedness, $\tau_D$ increases with $\epsilon$.

\item $\alpha_2(t)$ denotes the non-Gaussian parameter. It measures the deviation of the displacement distribution from a Gaussian form. It vanishes for ordinary Gaussian diffusion and becomes nonzero when the particle samples a heterogeneous mobility environment or experiences intermittent trapping.

\item $R(\epsilon)$ denotes the relative dispersion of the local hopping rate or local mobility in the one-dimensional calculation. It measures the amplitude of the local mobility fluctuations generated by the rugged site-energy landscape. In the one-dimensional analytical approximation, the short-time plateau of the non-Gaussian parameter is proportional to $R(\epsilon)$.

\item $G_s({\bf r},t|\rho_0)$ denotes the self-propagator, or self part of the van Hove function, for a specified initial distribution $\rho_0$. It is the probability distribution of particle displacements at time $t$.

\item $F_s(k,t)$ denotes the self-intermediate scattering function, obtained as the Fourier transform of the self-propagator. Its small-$k$ expansion contains the mean-square displacement and the fourth-order cumulant that determines $\alpha_2(t)$.

\item $\rho_0$ denotes the initial distribution of the particle over the rugged landscape. The present theory keeps this distribution explicit in order to describe finite-time preparation dependence.

\item $\beta_0$ and $q_0=\beta_0\epsilon$ specify the initial preparation. The case $\beta_0=0$ corresponds to uniform initial placement, $\beta_0=\beta$ corresponds to equilibrium preparation, $\beta_0>\beta$ corresponds to trap-biased preparation, and $\beta_0<0$ corresponds to high-energy-biased preparation.

\end{enumerate}

The key distinction is therefore the following. The ruggedness $\epsilon$ and energy-correlation length $\xi$ define the Gaussian energy landscape. This landscape generates a heterogeneous local mobility field with variance $C_D(0)$ and correlation length $\xi_D$. The function $C_D(t)$ is the time correlation of the local mobility sampled by the particle. Its normalized decay is the mobility-memory kernel. In the one-dimensional reduced notation, this kernel is written as $[1+a(\epsilon)\tau]^{-1/2}$, where $a(\epsilon)=2D_{\rm eff}^{1D}(\epsilon)/D_0$ is the dimensionless memory-decay rate, not a separate microscopic memory function.

%==========================================

\section*{Appendix B: Recovery of Gaussianity and the Fourth Cumulant}

\renewcommand{\theequation}{B\arabic{equation}}
\setcounter{equation}{0}

In this appendix we briefly discuss how the present theory is related to the familiar recovery of Gaussian diffusion in liquids and supercooled liquids. This discussion is useful because the long-time diffusion coefficient alone does not tell us how the full displacement distribution approaches its Gaussian form.

For a simple liquid, the self part of the van Hove function becomes nearly Gaussian after a relatively short time, once the particle displacement is produced by many weakly correlated molecular collisions. In a supercooled liquid, this recovery is delayed. Some particles remain trapped in cages while others undergo larger displacements. This dynamic heterogeneity produces a broad and non-Gaussian displacement distribution. The usual non-Gaussian parameter, $\alpha_2(t)$, measures the deviation of the self-displacement distribution from a Gaussian form, while the four-point susceptibility, often denoted $\chi_4(t)$, measures fluctuations in the dynamical mobility. In supercooled liquids, $\chi_4(t)$ often exhibits a pronounced peak, and the position of this peak shifts to longer times as the liquid is cooled. This behavior reflects the increasing lifetime and spatial extent of dynamic heterogeneity.

The rugged landscape problem considered in this paper has an analogous structure, but with an important difference. In a supercooled liquid, the heterogeneity is dynamically generated by the many-body motion. In the present problem, the heterogeneity is already present in the quenched landscape. Different spatial regions have different escape rates and therefore different effective mobilities. A tagged particle moving through the landscape samples this heterogeneous mobility field along its trajectory. At short and intermediate times, different trajectories may sample very different mobility histories. This produces non-Gaussian displacement statistics. At long times, if the particle samples many statistically independent regions of the landscape, a central-limit mechanism restores the Gaussian central part of the propagator.

The simplest way to see this is through the fourth moment. For an isotropic Gaussian displacement distribution in $d$ dimensions, the fourth moment is not independent of the second moment, but is fixed by
\begin{equation}
\langle r^4(t)\rangle_{\rm G}
=
\frac{d+2}{d}\,
\langle r^2(t)\rangle^2 .
\end{equation}
The excess fourth cumulant may therefore be defined as the deviation from this Gaussian value:
\begin{equation}
K_4(t)
=
\langle r^4(t)\rangle
-
\frac{d+2}{d}\,
\langle r^2(t)\rangle^2 .
\end{equation}
The non-Gaussian parameter is then the normalized form of this excess fourth cumulant:
\begin{equation}
\alpha_2(t)
=
\frac{d}{d+2}
\frac{\langle r^4(t)\rangle}{\langle r^2(t)\rangle^2}
-1
=
\frac{d}{d+2}
\frac{K_4(t)}{\langle r^2(t)\rangle^2}.
\end{equation}
Thus the recovery of Gaussianity does not require the absolute fourth cumulant $K_4(t)$ to vanish. It only requires that $K_4(t)$ grow more slowly than the Gaussian contribution $\langle r^2(t)\rangle^2$.

Thus the recovery of Gaussianity does not require the absolute fourth cumulant $K_4(t)$ to vanish. It only requires that $K_4(t)$ grow more slowly than the Gaussian contribution $\langle r^2(t)\rangle^2$. Since normal diffusion gives $\langle r^2(t)\rangle\sim t$, the Gaussian part of the fourth moment grows as $t^2$. Gaussianity is restored when the excess part grows more slowly than $t^2$.

In the local-mobility representation,
\begin{equation}
{\rm Var}\,A(t)
=
2\int_0^t d\tau\,
(t-\tau)C_D(\tau).
\end{equation}
where $C_D(\tau)$ is the time correlation of the mobility sampled by the particle.
This quantity is the single-particle analogue of a dynamic susceptibility: it measures fluctuations in the mobility history of the tagged particle. It is not identical to the conventional many-particle $\chi_4(t)$ of supercooled liquids, but it plays a closely related role.

If the spatial mobility correlations are short ranged and the particle samples the landscape diffusively, then the sampled mobility correlation decays asymptotically as

\begin{equation}
C_D(t)\sim t^{-d/2}.
\end{equation}

This immediately gives
\begin{equation}
{\rm Var},A(t)\sim t^{3/2},
\qquad d=1,
\end{equation}
\begin{equation}
{\rm Var},A(t)\sim t\ln t,
\qquad d=2,
\end{equation}
and
\begin{equation}
{\rm Var},A(t)\sim t,
\qquad d>2.
\end{equation}

Since the mean accumulated mobility grows linearly with time, $\langle A(t)\rangle\sim t$, the denominator in $\alpha_2(t)$ grows as $\langle A(t)\rangle^2\sim t^2$. The decay of $\alpha_2(t)$ is therefore obtained by dividing the growth law of ${\rm Var},A(t)$ by $t^2$. Thus, in one dimension, ${\rm Var},A(t)\sim t^{3/2}$ gives $\alpha_2(t)\sim t^{-1/2}$. In two dimensions, ${\rm Var},A(t)\sim t\ln t$ gives $\alpha_2(t)\sim(\ln t)/t$. For $d>2$, ${\rm Var},A(t)\sim t$ gives $\alpha_2(t)\sim t^{-1}$.\textit{ Hence the recovery of Gaussian diffusion is slowest in one dimension, marginal in two dimensions, and normal, $1/t$-like, in higher dimensions.}

This result gives a simple interpretation of the dimensional dependence. In three and higher dimensions, the excess fourth cumulant grows only linearly with time, while the Gaussian contribution to the fourth moment grows as $t^2$. The $t^2$ term therefore eventually overwhelms the non-Gaussian contribution, and the normalized fourth cumulant decays as $1/t$. This is the usual central-limit recovery of Gaussian diffusion. In two dimensions the recovery is marginal, because the excess fourth cumulant grows as $t\ln t$, giving the slower decay $(\ln t)/t$. In one dimension the recovery is much slower, because traps and barriers occur in series and cannot be bypassed; the excess fourth cumulant grows as $t^{3/2}$, and the normalized non-Gaussianity decays only as $t^{-1/2}$.

The analogy with supercooled liquids is therefore clear but not exact. In supercooled liquids, the peak in $\chi_4(t)$ reflects the growth and decay of many-particle dynamic heterogeneity. In the present quenched landscape problem, the corresponding heterogeneity is spatial and static, but the tagged particle converts it into a time-dependent mobility history. Non-Gaussianity is largest when different trajectories have sampled very different mobility environments. It decays only after the particle has sampled enough independent regions for central-limit averaging to dominate. The fourth cumulant, and especially its normalized form $\alpha_2(t)$, therefore provides the simplest measure of the approach to Gaussianity in rugged energy landscapes.

%=====================  BIBLIOGRAPHY ============================

\end{document}